\documentclass[10pt,letterpaper]{article}
\usepackage{graphicx}
\usepackage{amssymb,amsmath}
\usepackage{float}
\usepackage{lipsum} 

\usepackage[sc]{mathpazo} 
\usepackage[T1]{fontenc} 
\usepackage{microtype} 
\usepackage[hmarginratio=1:1,top=32mm,columnsep=20pt]{geometry} 

\usepackage[hang, small,labelfont=bf,up,textfont=it,up]{caption} 
\usepackage{booktabs} 
\usepackage{float} 
\usepackage{hyperref} 

\usepackage{lettrine} 
\usepackage{paralist} 

\usepackage{abstract} 

\usepackage{titlesec} 
\renewcommand\thesection{\Roman{section}} 
\renewcommand\thesubsection{\Roman{subsection}} 
\titleformat{\section}[block]{\large\scshape\centering}{\thesection.}{1em}{} 
\titleformat{\subsection}[block]{\large}{\thesubsection.}{1em}{} 

\usepackage{fancyhdr} 

\title{\vspace{-15mm}\fontsize{24pt}{10pt}\selectfont\textbf{Comprehensive Numerical Modelling of a Low-Gain Optical Parametric Amplifier as a Front-End Contrast Enhancement Unit}} 

\author{
\large
\textsc{A. B. Sharba, G. Nersisyan, M. Zepf,  M. Borghesi and  G. Sarri}\\
\normalsize Centre for Plasma Physics, School of mathematics and Physics, Queen's University Belfast \\ 
\normalsize \href{mailto:asharba01@qub.ac.uk}{asharba01@qub.ac.uk} 
\vspace{-5mm}
}
\date{}

\begin{document}

\maketitle 


\begin{abstract}

We present a comprehensive model for predicting the full performance of a second harmonic generation-optical parametric amplification system that aims at enhancing the temporal contrast of laser pulses. The model simultaneously takes into account all the main parameters at play in the system such as the group velocity mismatch, the beam divergence, the spectral content, the pump depletion, and the length of the nonlinear crystals. We monitor the influence of the initial parameters of the input pulse and the interdependence of the two related non-linear processes on the performance of the system and show its optimum configuration. The influence of the initial beam divergence on the spectral and the temporal characteristics of the generated pulse is discussed. In addition, we show that using a crystal slightly longer than the optimum length and introducing small delay between the seed and the pump ensures maximum efficiency and compensates for the spectral shift in the optical parametric amplification stage in case of chirped input pulse.  As an example, calculations for bandwidth transform limited and chirped pulses of sub-picosecond duration in beta barium borate crystal are presented.

\end{abstract}


\section{Introduction}

The development of chirp pulse amplification (CPA)techniques \cite{strickland1985compression} has enabled exciting and rapidly developing investigations of laser-matter interactions at relativistic intensities. In order to ensure a clean interaction of the high intensity peak of the laser pulse with the target, it is necessary to avoid deformation or heating induced by lower-intensity pedestals or pre-pulses. For instance, a non-ideal laser contrast dramatically affects laser-driven ion acceleration, imposing stringent limits on the thickness of the target \cite{x}. Thus, generating laser pulses that are devoid of any pre-pulses or pedestal is one of the key elements in the development of the ultra-high-intensity laser systems .

The temporal contrast of high intensity lasers has been enhanced using many techniques, such as cross polarized wave generation technique (XPW) \cite{chvykov2006generation}, nonlinear ellipse rotation \cite{homoelle2002pulse, zhang2008pulse}, saturable absorber \cite{liu2010contrast}, double chirped pulse amplification \cite{kalashnikov2005double} and optical parametric amplification (OPA) \cite{shah2009high}. Furthermore, plasma mirror techniques \cite{gold1994direct, dromey2004plasma} can be used after the main compressor to enhance the contrast by four orders of magnitude using two plasma mirrors, albeit at the cost of reducing the laser energy. OPA has advantages over the other pulse cleaning techniques since it is a second
order nonlinear process which does not require high pulse intensities, and therefore effectively minimizing the high order effects. Furthermore,
OPAs have many attractive features such as broad amplification bandwidth, high gain and no amplified spontaneous amplification accumulation.

In particular, different configurations \cite{shah2009high, jovanovic2006optical,huang2011ultrashort} have been proposed in order to enhance the temporal contrast of laser pulses in the infrared range employing OPA. Short pulse low gain OPA is one of the simplest and efficient techniques which can work at the front end of high-power laser systems. It relies on the fact that the gain in the OPA occurs only during the pump pulse duration. Thus, using short pump pulse, can ensure that the pump will match only the main pulse of the seed without amplifying anything outside it. This will lead to the generation of an idler pulse of extremely high contrast, since it is generated only during the optical parametric interaction \cite{shah2009high}. 

Short-pulse and low gain OPAs, as temporal contrast enhancement systems, have received a significant degree of interest, because, besides the efficient contrast enhancement ability, they can be integrated into an existing high power laser system without imposing major changes to the system \cite{shah2009high}. However, in short pulse low gain regime, small difference in the used crystal thickness or/and slight different system configuration can result in generation of degraded pulse or simply limiting the unit efficiency, as will be shown later. Thus, careful characterization and optimization of the unit is of a particular importance in order to seed a high power laser system.

\begin{figure}[h!]
\centering\includegraphics[ height=5.5cm, width=10cm]{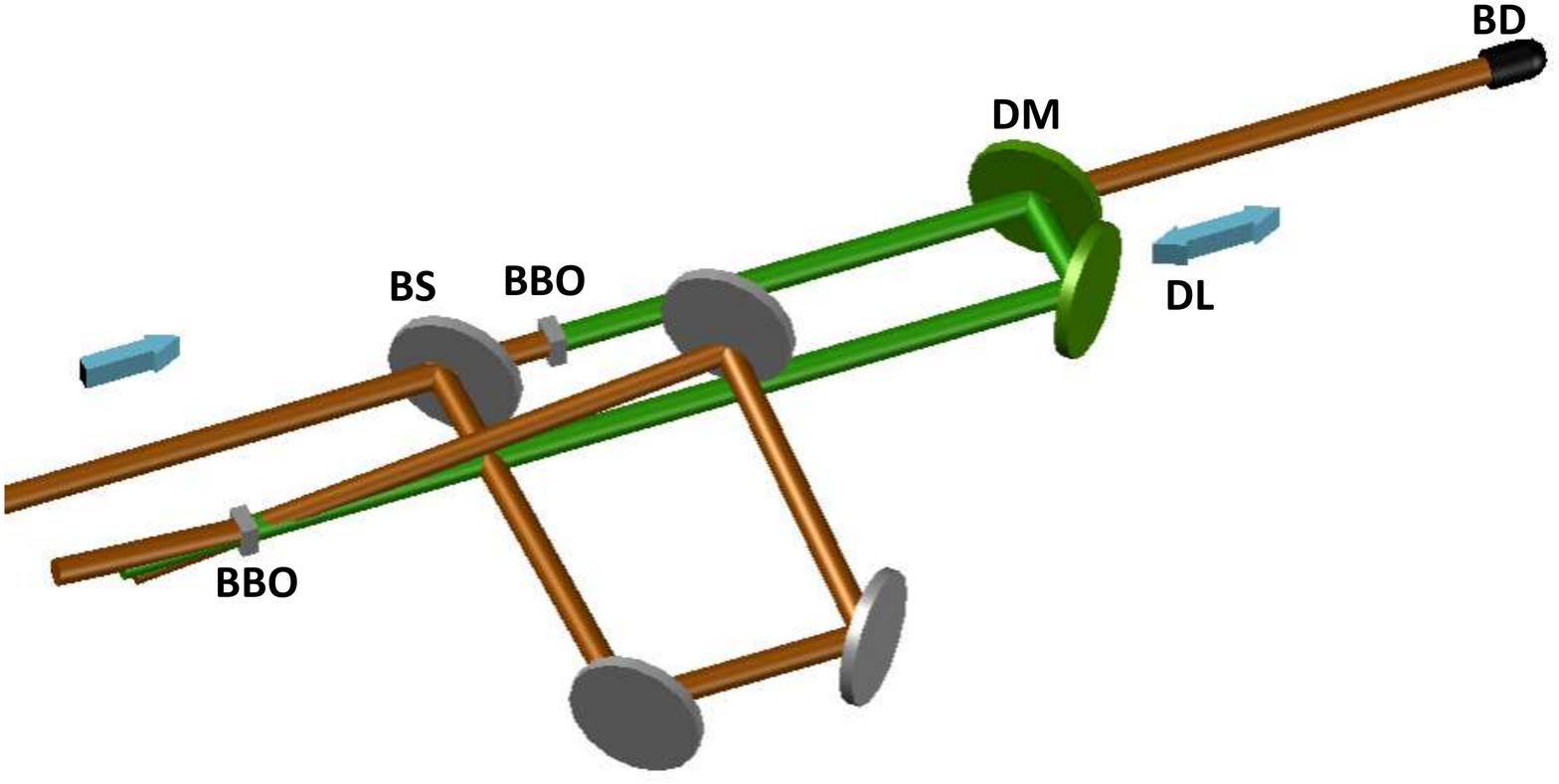}
\caption{A layout of a SHG-OPA based temporal contrast enhancement system, BS: beam splitter, DM: dielectric mirror, BD: beam damper and DL: delay line.}
\label{fig1}
\end{figure}

In its simplest configuration, we will consider here a system, similar to that in \cite{shah2009high}, comprising a second harmonic generation (SHG) stage and an OPA stage. The input beam is split into two parts, the large part is frequency doubled to generated the pump pulse and the seed of the OPA will be the smaller part of the input beam, as shown in figure(\ref{fig1}). The idler pulse is considered as the useful output of the system due to the high contrast of this pulse.

 Although the SHG and the OPA processes, individually, have been extensively studied over the years, a comprehensive detailed study that simultaneously considering all the main parameters at play and investigating how the first process influences the second in a system which is driven by a single source, has not yet been carried out according to our knowledge. In such a system, the seed energy is determined only by the splitting ratio of the beam splitter, as shown in figure (\ref{fig1}), while the efficiency of the SHG process is another factor that determines the pump energy. Therefore, since the OPA works in low gain regime, the SHG efficiency has significant impact on the OPA, in terms of saturation effects. Furthermore, the duration and the temporal shape of the pump pulse, which are set in the SHG stage, define the shape of the idler pulse and set the time window of the gain and the superfluorescence in the OPA process. At optimum performance of the SHG, the resulting pulse, which will pump the following OPA, could be shorter than the input pulse due to the quadratic dependence of the frequency conversion process. This initial parameter makes the OPA work in an environment that is different than that of the traditional OPAs which are pumped by pulses which are longer or equal to the seed pulse duration. Accordingly, characterizing the two related nonlinear processes in linked mode is of particular importance.  
 
 The effect of beam divergence on the nonlinear interaction has been studied by many authors since the invention of the laser. Early studies \cite{kleinman1962theory, bjorkholm1966optical}, focused on low conversion efficiency where the effect could be investigated analytically \cite{wong1992beam}. S. K. Wong, et al \cite{wong1992beam} studied the effect of beam divergence on the frequency conversion efficiency for laser intensity up to 2 GW/cm$^{2}$. In their work they assumed constant field amplitude in the temporal domain and the group velocity mismatch was not taken into account, while, as will be shown later, the influence of beam divergence is different for different pulse durations. The study was extended by I.Jovanovic, et al \cite{jovanovic2001angular} to predict the performance of OPAs, of pump intensity up to 1 GW/cm$^{2}$, taking into account the beam quality and assuming long pulse duration where the group velocity mismatch can be ignored. 

On the other hand, the effect of group velocity mismatch (GVM) on the SHG conversion efficiency and the temporal profile of laser pulses was reported by many authors, for example in \cite{krylov1995second,zhang1998second}. However, they considered the laser beam as a plane wave, where all the beam parts satisfy perfect phase matching. A numerical study has been carried out by A. Dement’ev, et al \cite{dement2004numerical} to simulate the performance of short pulse OPA taking into account group velocity mismatch, diffraction and other parameters. However, in their study the pump pulse duration was considered much longer than that of the seed. Several other works \cite{zhang2001simulation, wang2003efficiency} have included diffraction, the GVM and dispersion effects in the calculations to simulate the nonlinear frequency conversion processes. A numerical study for simulating SHG-OPA based temporal contrast cleaner has been done by Y. Wang \cite{wang1994optical} for 1 ps input pulse. The simulation is for type II SHG and OPA, assuming plane-wave approximation. The divergence of the beam, the spectral content and chirp rate of the input pulse were not considered in their simulation.     
 
 In this work we present a comprehensive simulation of a SHG-OPA temporal cleaning unit that predicts the performance of the unit for any initial conditions within the range of pulse duration and input intensity where the phase modulations and the high order dispersion can be ignored. The model simultaneously takes into account the temporal walk-off of the interacting fields due to the GVM, the effects of the divergence of the beams and the resulting reconversion processes, the spectral content of the pulses, the pump depletion, and the length of the nonlinear crystal, which are the most effective parameters in this regime. Considering these parameters simultaneously shows their interdependence and sets an effective range for each of them. We show results for a system driven by a single input pulse, to be temporally cleaned. Equally important, we explore the influence of these parameters on the interdependence of the related non-linear processes in the system. We show how the effect of the beam divergence depends on the input pulse duration and how this source of phase mismatch can influence the temporal and the spectral profiles of the resulting pulses. Also, we present a detailed investigation on the effect of the pulse chirp and the spectral content of the input pulse on the performance of the system. By considering all these effects, we present an optimum configuration for the system in order to generate a pulse with temporal, spectral and spatial characteristics that are suitable for seeding a high power laser system. We include results for transform limited (TL) pulses, which are generated from ideal systems, and for non-transform limited (non-TL) pulses which can be a result of imperfect pulse compression in the generating system. 

In this letter, the most relevant theoretical concepts and the assumptions of the model are listed in section II. The two parts of section III are devoted to the results of the model and the discussion of the SHG and the OPA stages respectively. Finally, the conclusions of the work are drawn in section IV.    


\section{Numerical Simulations}

When a light wave at the fundamental frequency (FF) enters a non-linear medium, the complex amplitudes of the FF and the SH can be estimated by solving a system of  coupled differential equations which, in the slowly varying amplitude approximation, can be given by: \cite{krylov1995second}:
\begin{subequations}
\begin{equation}
\frac{\partial A_{1}}{\partial z}+\frac{1}{v_{g1}}\frac{\partial A_{1}}{\partial t}=-i\frac{2\omega d_{eff}}{n_{1}c}A_{1 }^{*}A_{2 } \exp(-i\Delta kz)
\label{eq1}
\end{equation}

\begin{equation}
\frac{\partial A_{2}}{\partial z}+\frac{1}{v_{g2}}\frac{\partial A_{2}}{\partial t}=-i\frac{2\omega d_{eff}}{n_{2}c}A_{1 }^{2} \exp(i\Delta kz)
\label{eq2}
\end{equation}
\label{eqsd2}
\end{subequations}
Where $A_{j} , n_{j}$ and $v_{gj}, j=1,2$ are the field amplitude, the refractive index, and the group velocity at the FF and the SH frequencies respectively, $ \omega $ the fundamental angular frequency, $d_{eff}$ is the effective nonlinear coefficient, $\Delta k$ is the phase mismatch, z is the propagation distance of the waves along the crystal and $c$ is the speed of light.

When the input pulse duration is longer than 100 fs, the second and higher order dispersions can be neglected, where the dispersive length of the pulse in this case is much longer than the used crystals. Also, within the intensity level of the working environment of the low gain temporal contrast enhancement system, the self and the cross phase modulations are not related.

By transforming the time frame of equations (\ref{eqsd2}) to the frame of an observer moving with the higher frequency pulse $\tau =t-z/v_{g2}$, we obtain \cite{cerullo2003ultrafast}:

\begin{subequations}
\begin{equation}
\frac{\partial A_{1}}{\partial z}+(\frac{1}{v_{g1}}-\frac{1}{v_{g2}})\frac{\partial A_{1}}{\partial \tau}=-i\frac{2\omega d_{eff}}{n_{1}c}A_{1 }^{*}A_{2} \exp(-i\Delta kz)
\label{eq3}
\end{equation}

\begin{equation}
\frac{\partial A_{2}}{\partial z}=-i\frac{2\omega d_{eff}}{n_{2}c}A_{1}^{2} \exp(i\Delta kz)
\label{eq4}
\end{equation}
\end{subequations}

In order to introduce the divergence of the beam into the solution, we apply the assumption of S. K. Wong, et al \cite{wong1992beam} by considering the beam as a summation of many plane waves that travel at slightly different angles with respect to the propagation direction. The central plane-wave component enters the crystal at the phase matching angle while the others have small deviation angles. The distribution function of the divergent components and the corresponding wave vector mismatch can be found in the same reference. In this case, the FF and the SH fields amplitudes can be estimated from:

\begin{subequations}
\begin{equation}
\frac{\partial A_{1 j}}{\partial z}+(\frac{1}{v_{g1}}-\frac{1}{v_{g2}})\frac{\partial A_{1 j}}{\partial \tau}=\sum_{k=1}^{N}-i\frac{2\omega d_{eff}}{n_{1}c}A_{1 k }^{*}A_{2 kj } \exp(-i\Delta k_{kj}z)
\label{eq3}
\end{equation}

\begin{center}
$j=1, 2, 3, ..... N$
\end{center}

\begin{equation}
\frac{\partial A_{2 kj}}{\partial z}=-i\frac{2\omega d_{eff}}{n_{2}c}A_{1 k}A_{1 j} \exp(i\Delta k_{kj}z)
\label{eq4}
\end{equation}
\begin{center}
$j,k=1, 2, 3, ..... N$
\end{center}
\end{subequations}
The field amplitude of any of the interacting fields is given by:
\begin{equation}
A=\left [ \sum_{j=1}^{N} A_{j}A_{j}^{*} \right ]^{1/2}
\end{equation}

Similarly for OPA:

\begin{subequations}
\begin{equation}
\frac{\partial A_{sj}}{\partial z}+(\frac{1}{v_{gs}}-\frac{1}{v_{gp}})\frac{\partial A_{sj}}{\partial \tau}=\sum_{k=1}^{N}-i\frac{\omega_{s}d_{eff}}{n_{s}c}A_{ikj }^{*}A_{pk } \exp(-i\Delta k_{kj}z)
\label{eq1}
\end{equation}
\begin{center}
$j=1, 2, 3, ..... N$
\end{center} 
\begin{equation}
\frac{\partial A_{ikj}}{\partial z}+(\frac{1}{v_{gi}}-\frac{1}{v_{gp}})\frac{\partial A_{ikj}}{\partial \tau}=-i\frac{\omega_{i}d_{eff}}{n_{i}c}A_{sj}^{*}A_{p k} \exp(-i\Delta k_{kj}z)
\label{eq1}
\end{equation}

\begin{center}
$j,k=1, 2, 3, ..... N$
\end{center}

\begin{equation}
\frac{\partial A_{pj}}{\partial z}+\frac{\partial A_{pj}}{\partial \tau}=\sum_{k=1}^{N}-i\frac{\omega_{p}d_{eff}}{n_{p}c}A_{s k}A_{ikj } \exp(i\Delta k_{kj}z)
\label{eq1}
\end{equation}
\begin{center}
$j=1, 2, 3, ..... N$
\end{center} 
\end{subequations}

where, $ s,$$\: i$$ \:$ and $p$ denote seed, idler and pump respectively.

It is worth to note that the Kerr-like nonlinearity \cite{bache2013anisotropic} is not likely to change the characteristics of the  pulse in the range of the study. In fact, the phase mismatch due to the beam divergence is not large enough to give rise to a sufficient number cycles of up and down conversion processes within the pulse splitting length that can influence the phase of the generated pulses.

The coupled equations are solved numerically when the group velocity and the phase matching structures are considered, e.g. for a crystal of thickness comparable to or longer than the pulse splitting length L$_{s}$, which is the  minimum distance after which the pulses walk away from each other  \cite{nautiyal2009effects}
\begin{equation}
L_{s}=\left | \frac{1}{v_{g1 }}-\frac{1}{v_{g2}} \right |^{-1} T
\label{eqx}
\end{equation}
Where, $T$ is the pulse duration at the full width half maximum (FWHM). In case of OPA, GVM is the difference between the group velocity of the pump and the projection of the signal group velocity on the pump direction , $ v_{g1 } \cos \Omega $, where $ \Omega $ is the non-collinear angle.

We use the quadratic spectral broadening $f$ to represent the chirp rate of the pulse. $f$ equals zero for TL pulses and is positive for chirped pulses.  For a Gaussian pulse $f$ can be defined by \cite{zhang1998second}:  
\begin{equation}
T=\frac{4 \ln2}{\Delta \omega}({1+(4f)^{2}})^{1/2}
\label{eq1}
\end{equation}
Where, $\Delta \omega$ is the angular frequency bandwidth of the pulse at the FWHM.

Pump depletion, GVM and beam divergence will all affect the spatial profiles of the interacting beams as they propagate along the crystal. Thus, in order to monitor the beam quality as the beam propagates into the crystal, the quality factor M$^{2}$ can be calculated from \cite{dement2002numerical} 
\begin{equation}
M^{2}(t)=\frac{\left \{ \int_{0}^{\infty}\left | \frac{\partial E}{\partial r} \right |^{2} rdr\int_{0}^{\infty}\left | E \right |^{2} r^{3}dr-\frac{1}{4}\left | \int_{0}^{\infty}r^{2}\left [ \frac{\partial E}{\partial r}E^{*}-\frac{\partial E^{*}}{\partial r}E \right ]dr \right |^{2}\right \}^{1/2}}{\int_{0}^{\infty}\left | E \right |^{2} rdr}
\label{eqxx1}
\end{equation}
Where $E$ is the normalized transverse electric field amplitude distribution, and $r$ is the transverse dimension.

The quality factor of the entire pulse is, then, the power-weighted time average quality parameter \cite{dement2002numerical}.
\begin{equation}
\left \langle M^{2} \right \rangle=\frac{\int_{-\infty}^{\infty}M^{2}(t)P(t)dt}{\int_{-\infty}^{\infty}P(t)dt}
\label{eqxxs}
\end{equation}

We solved the nonlinear coupled equations numerically using the fourth-order Range-Kutta method \cite{blum1962modification}. The input beam is considered as a perfect Gaussian beam in the temporal and the spatial domains. The simulation code consists of three parts: in the first part the optical properties of the crystal and the distribution of the input pulse in time and space are calculated along with the angular distribution of the divergent components. The second part is devoted to the numerical integrations of the coupled equations. Finally, the energies and the intensities of the interacting pulses, the conversion efficiency of the process and full characterization of the resulting pulses in time, space, and spectral domains are found in the third part. After every step of the integration, at every slice of the crystal, the temporal shapes of the interacting fields are determined and then fed to the next integration step.  The simulation codes can be used for any initial parameters of the input pulse and the system in the range where phase modulations and dispersion can be ignored.

The accuracy of the numerical solution is determined, first, by checking the deviation of the solution when changing the integration step size and the number of the plane-wave components, second, by verifying the total energy conversion efficiency of the interacting pulses along the crystal. Finally, the accuracy of the whole simulation is verified by comparing the results of the simulation with the known solutions in cases of long pulse duration, monochromatic wave and plane-wave approximations.


\section{Results}

According to our initial assumption, we will focus our attention here on pulses of sub-picosecond duration and of intensity in the range 1-10 GW/cm$^{2}$ in Beta Barium Borate (BBO) crystals ($d_{eff}=2.2$ pm/V) of different thicknesses. The input beam spectrum is considered to be Gaussian of central wavelength 1053 nm. The calculations include results of TL and linearly chirped pulses. Since, in practice, the crystal length is not a parameter that can be easily fine tuned to give optimum result, we show results of the nonlinear interactions for distances that extend beyond the optimum crystal length. Most of the results are for 500 fs pulses of spectral bandwidth $\Delta \lambda\approx$ 3.26, 3.5, 4.2 and 6.2 nm corresponding to values of the quadratic spectral broadening of $f$ = 0, 0.1, 0.2, and 0.4 respectively.

\subsection{Second Harmonic Generation}
\subsubsection{The effect of beam divergence on the conversion efficiency in short pulse regime}
When a divergent beam of long pulse duration enters a nonlinear medium it experiences sequential up and down frequency conversion processes due to the phase mismatch of the paraxial parts of the beam. Thus, the efficiency of the conversion process has an oscillatory behaviour with frequency defined by the interplay of the beam divergence and the input intensity. High input intensity or/and large beam divergence result in a high oscillation frequency \cite{wong1992beam}. However, in the case of short pulses, the oscillation is damped due to the temporal walk-off of the interacting pulses, which causes changes in  the local overlapped-intensities as the pulses propagate into the nonlinear medium, leading to a derailing in the sequence of the non-linear processes.

In order to show the dependence of the effect of the beam divergence, or any source of phase mismatching, on the duration of the input pulse, it is more convenient first to find a representation for the coherence length that takes into account pump depletion. For this purpose, we define a characteristic length $L _{m} ( \Delta  k,I _{\omega o}) $, which is, in long pulse duration regime, the crystal length after which the SHG process changes its direction from up-conversion to down-conversion, i.e. the length for the first efficiency maxima.

\begin{equation}
L_{m}=\frac{1}{4}\left [ -L_{NL}^{2}\Delta k  K(\gamma ) +L_{NL}\sqrt{16+L_{NL}^{2}\Delta k^{2}}  K(\gamma )\right ]
\label{eqxxs}
\end{equation}

Where, $K(\gamma)$ is the complete elliptic integral of $\gamma$

\begin{equation}
\gamma= \left ( -\frac{L_{NL}\Delta k}{4}+\sqrt{1+\frac{L_{NL}^{2}\Delta k^{2}}{16}} \right )^{2} 
\label{eqxxs}
\end{equation}

\begin{equation}
L_{NL}=\frac{c}{2\omega d_{eff}}\sqrt{\frac{2\varepsilon _{o}n_{\omega}^{2}n_{2\omega }c}{I_{\omega }}}
\label{eqxxs}
\end{equation}

\begin{figure}[h!]
\centering\includegraphics[height=6cm, width=9cm]{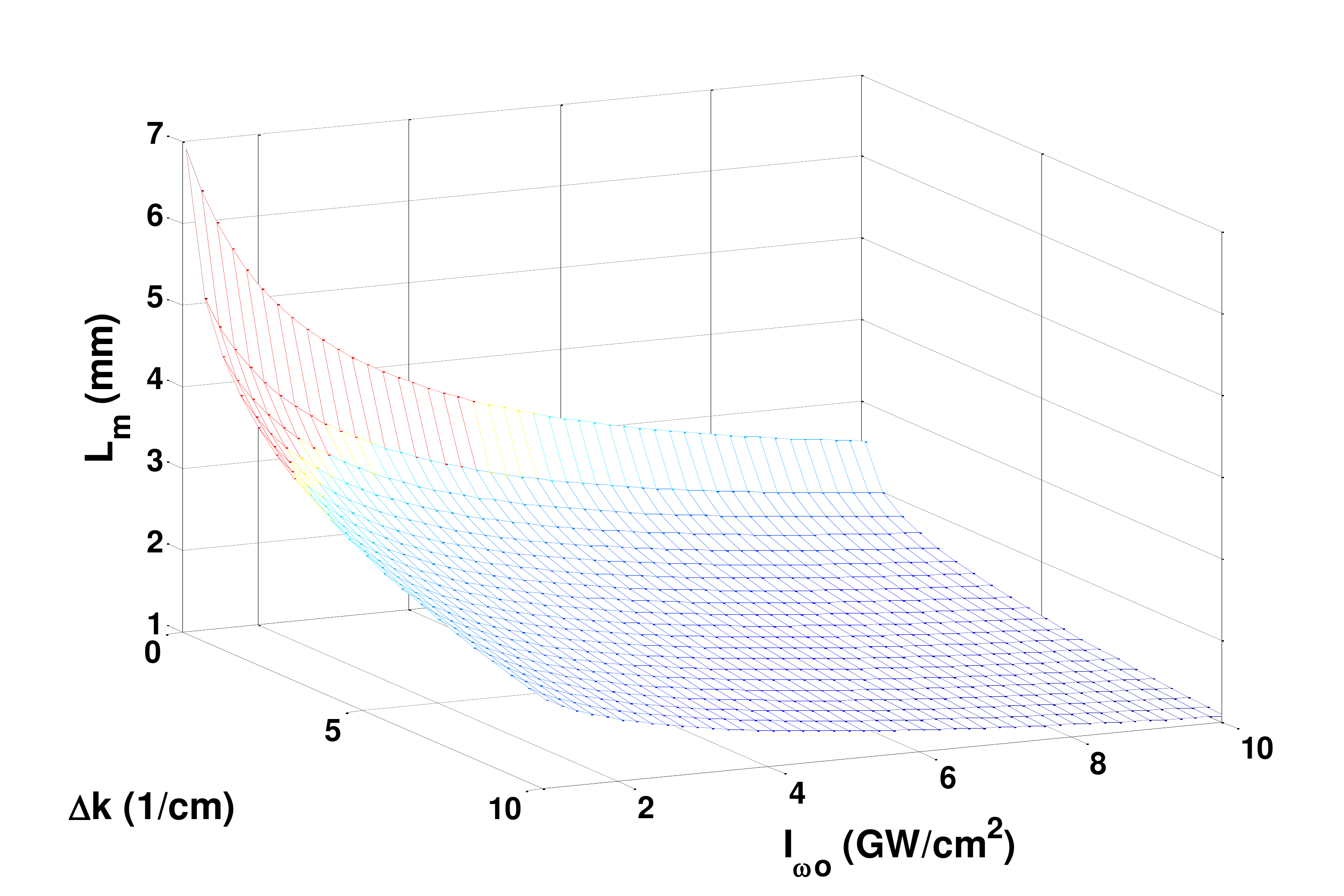}
\caption{The dependence of $L_{m}$ on the initial phase mismatch and the input intensity in BBO crystal. }
\label{fig2}
\end{figure}

In fact, $ L_{m}$ represents the interplay of the phase mismatch and pump depletion. Its dependence on $\Delta k$ and the input intensity is shown in figure (\ref{fig2}).

Accordingly, for short pulses, the effect of beam divergence, or any source of phase mismatching, on the conversion efficiency depends on the ratio between L$_{s}$ and $L_{m}$. For pulses of L$_{s}$ that is shorter than $L_{m}$ the divergence of the beam only limits the maximum achievable efficiency and no oscillation can be seen in the nonlinear interaction. This happens because the pulses walk away from each other before the beginning of the back conversion process. However, the down conversion process starts taking place and reduces the SHG intensity as the ratio between the L$_{s}$ and $L_{m}$ increases. 

\begin{figure}[h!]
\centering\includegraphics[height=9cm, width=11cm]{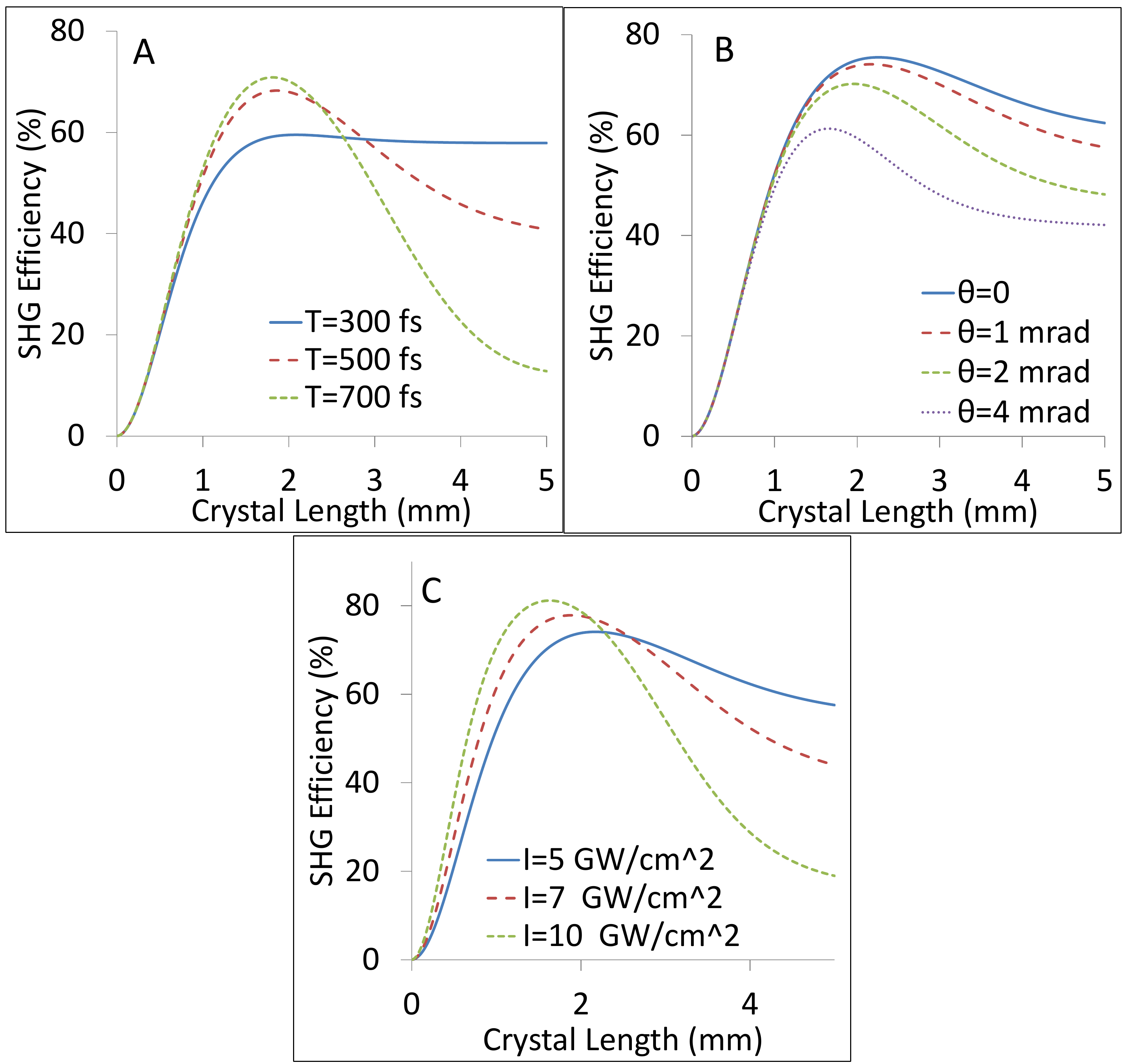}
\caption{SHG efficiency as a function of the crystal length of (A) pulses of  5 GW/cm$^{2}$ intensity, 1 mrad divergence, 5.44 nm bandwidth and of different durations, (B) TL pulses of 5 GW/cm$^{2}$ intensity,500 fs duration, about 3.26 nm bandwidth and different beam divergence angles, (C) TL pulses of 500 fs duration and 1 mrad beam divergence, about 3.26 nm bandwidth and different input intensities. }
\label{fig3}
\end{figure}

Figure(\ref{fig3}) presents three examples of the SHG conversion efficiency behaviour as a function of the crystal length. Frame (A) of the figure shows the conversion efficiency of pulses of similar divergence 1 mrad, intensity 5 GW/cm$ ^{2} $ and bandwidth 5.44 nm but having different durations, i.e. different L$_{s}$. The effect of the reconversion emerges when the pulses can stay overlapped for longer distance. In this case, the increment of the maximum achievable  efficiency is due to the reduction of the influence of the temporal walk-off of the interacting fields over the interaction length as the input pulse becomes longer. Frame (B) of the figure illustrates the effect of shortening  the coherence length, i.e. increasing the beam divergence, on the conversion efficiency of TL pulses of the same duration and intensity. The significance of the back-conversion process rises up as the divergence of the input beam increases. However, the down-conversion process is terminated gradually as the pulses walk away from each other. The curves in frame (C) of the figure are for TL pulses of the same duration and of the same beam divergence, but with different input intensities. As shown, the same divergence of the input beam can result in stronger reconversion process at high input intensity due to the  shortening of depletion length. This means, at high input intensity, the optimum crystal length is very critical.

In \cite{eimerl1987high} D. Eimerl concluded that demagnifying a beam of a particular peak power does not enhance the conversion efficiency, but only reduces the required crystal length for maximum efficiency. However, for short pulses the pulse splitting length is an additional parameter that should be considered, because telescoping a short pulse is an imperative in order to reach the maximum conversion efficiency within the pulse splitting length. In this case, the increment of the divergence due to demagnifying the beam has unavoidable effects on the conversion efficiency. 

\subsubsection{The influence of the chirp rate of the input pulse on the conversion efficiency}
\begin{figure}[h!]
\centering\includegraphics[ height=5cm, width=11cm]{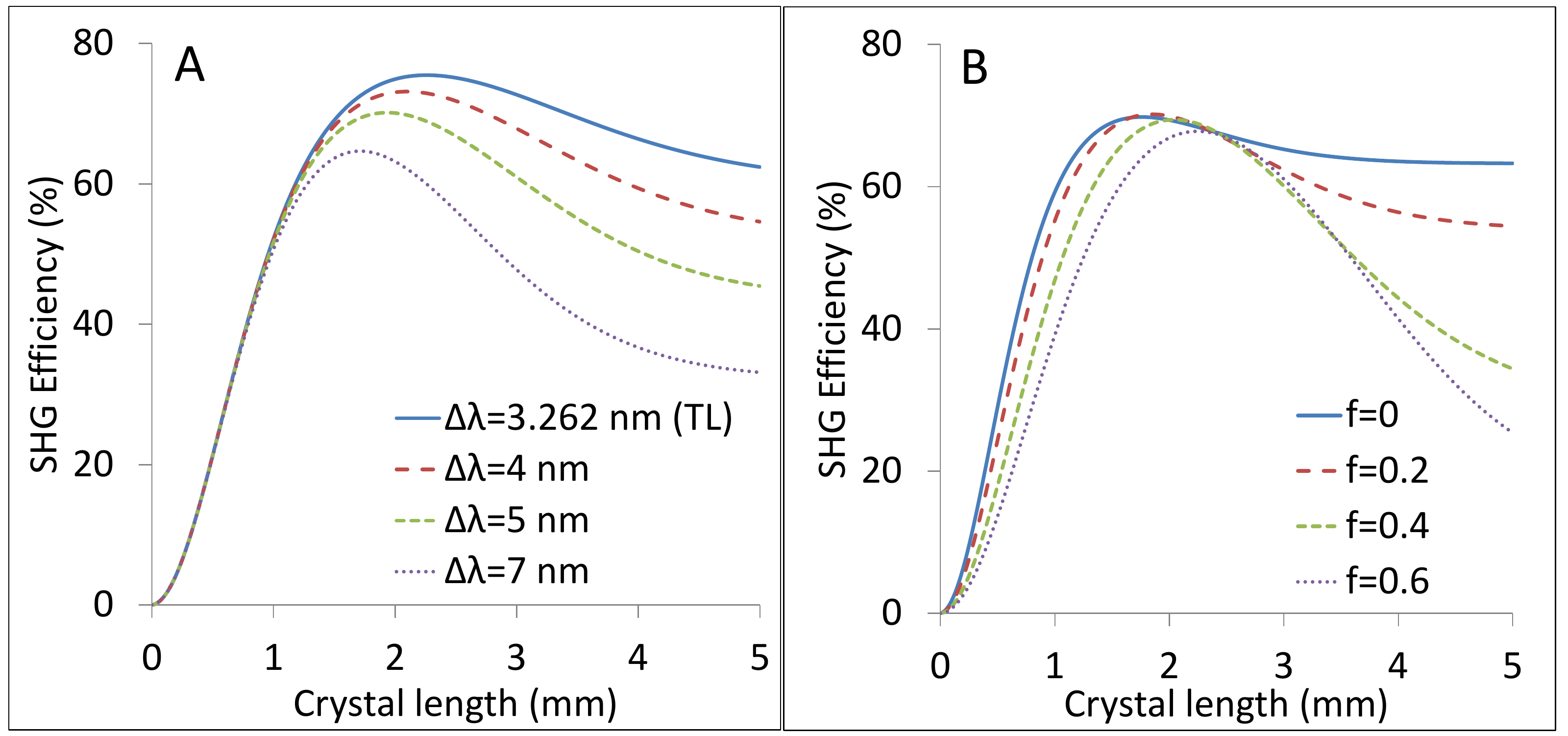}
\caption{SHG efficiency as a function of the crystal length of (A) pulses with 500 fs duration, 5 GW/cm$ ^{2} $ intensity and different bandwidths, (B) pulses of 5 nm spectral bandwidth and different chirp rates, $f$ values correspond to $T$= 326, 418, 615 and 850 fs respectively, the input beam having energy 170 $\mu$J and 2.8 mm diameter,  .}
\label{fig4}
\end{figure}

The spectral content of the input pulse is another inherent source of phase mismatch that controls the efficiency of the frequency conversion process of short pulses. Figure(\ref{fig4} A) shows the SHG conversion efficiencies as function of the crystal length for pulses of the same duration and input intensity but different spectral bandwidths. A broader spectral bandwidth of the input pulse implies that the amount of the pulse energy that deviates from the perfect phase matching increases, limiting the maximum achievable efficiency and causing the down conversion process to take place. As an effect, there is a similar dependence on the duration and on the intensity of the input pulse as for the beam divergence. 

On the other hand, for a particular laser system that generates pulses of given energy and bandwidth, broadening of the pulse duration, arising due to imperfect compression or any other reason, increases the crystal length required  for the maximum conversion efficiency, as shown in figure(\ref{fig4} B). In addition, for longer input pulses the optimum crystal length is more critical, because broadening of the pulse provides sufficient length for the back conversion process to take place.  

\subsubsection{The influence of the spectral content of the input pulse on the temporal profile of the SHG pulse}
\begin{figure}[h!]
\centering\includegraphics[ height=5cm, width=6cm]{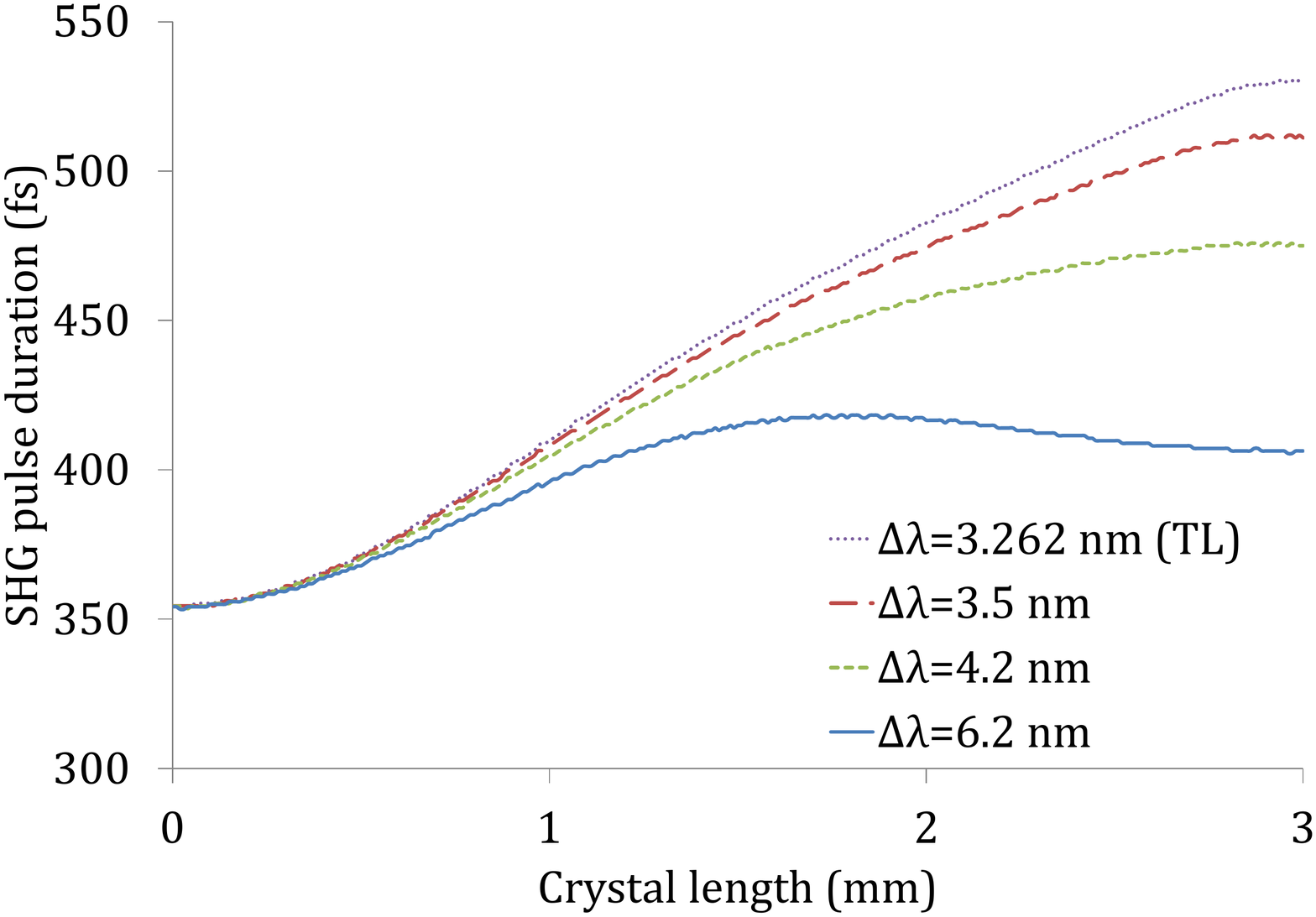}
\caption{The variation of the duration of the SHG pulses that are generated from input pulses of different spectral bandwidths all of 500 fs duration and intensity 5 GW/cm$^{2}$ as a function of the crystal length.}
\label{fig5}
\end{figure}

As to what concerns the temporal duration, the bandwidth of the input pulse determines also the duration and the temporal shape of the resulting SHG pulse. At the same input pulse duration, a shorter SH pulse can be obtained from an input pulse of broader bandwidth. This is because, for chirped pulses, the spectral components that are at the leading front and the tail of the pulse, i.e. at the wings of the spectrum,  undergo low conversion efficiency due to the deviation of these components from the perfect phase matching. Damping these components leads to shortening the resulting SHG pulse. The variation of the duration of the SHG pulse along the crystal is plotted in figure (\ref{fig5}). The pulses in the figure are generated from 500 fs input pulses, all with intensity 5 GW/cm$^{2}$ but having different bandwidths.
 
As shown, over the first slice of the crystal progressive broadening in the pulse duration is taking place due to the GVM between the interacting pulses, after which, for non-TL pulses, the back-conversion process starts competing with the broadening of the pulse. The influence of the reconversion process in reducing the pulse duration becomes more pronounced for broader bandwidth of the input pulse. In fact, this shortening in the generated pulse duration corresponds to a broadening in the TL duration of the pulse due to the reduction in the spectral bandwidth.  
\subsubsection{The impact of the input intensity on the pulse temporal profile}
\begin{figure}[h!]
\centering
\includegraphics[ height=5cm, width=10cm]{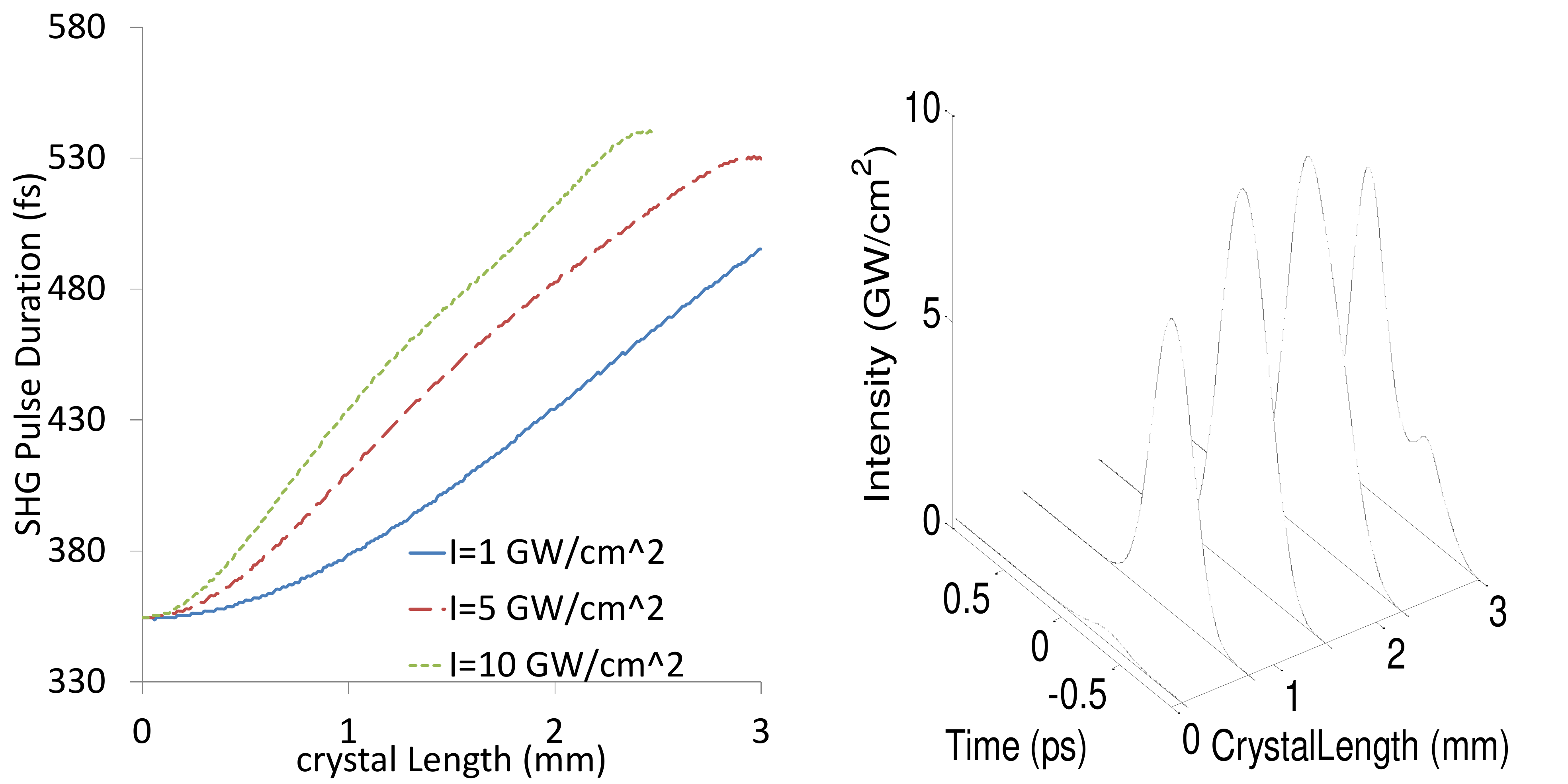}
\caption{A: SHG pulse duration as a function of the crystal length at  different pump intensities, input pulse duration is 500 fs, B: temporal profiles of SHG pulse that is resulting from 500 fs pulse of 10 GW/cm$^{2}$ intensity at different places along the crystal .}
\label{fig6}
\end{figure}

Another initial parameter that affects the temporal shape of the resulting SHG pulse is the intensity of the input pulse. At high input intensity GVM causes more pulse broadening due to the high energy exchange level. Figure(\ref{fig6}A) shows the broadening of SHG pulses that are generated from TL pulses of different input intensities. The curve at high input intensity is drawn up to a certain point in the crystal after which the SHG pulse start splitting into multiple peaks due to the back conversion process, as shown in part (B) of the figure. The crystal length after which the SHG pulse is modified depends on the interplay of the initial phase mismatch and the input intensity.     

\subsubsection{The dependence of the spectral and the temporal profiles of the SHG pulse on the beam divergence}
\begin{figure}[h!]
\centering
\includegraphics[ height=7cm, width=10cm]{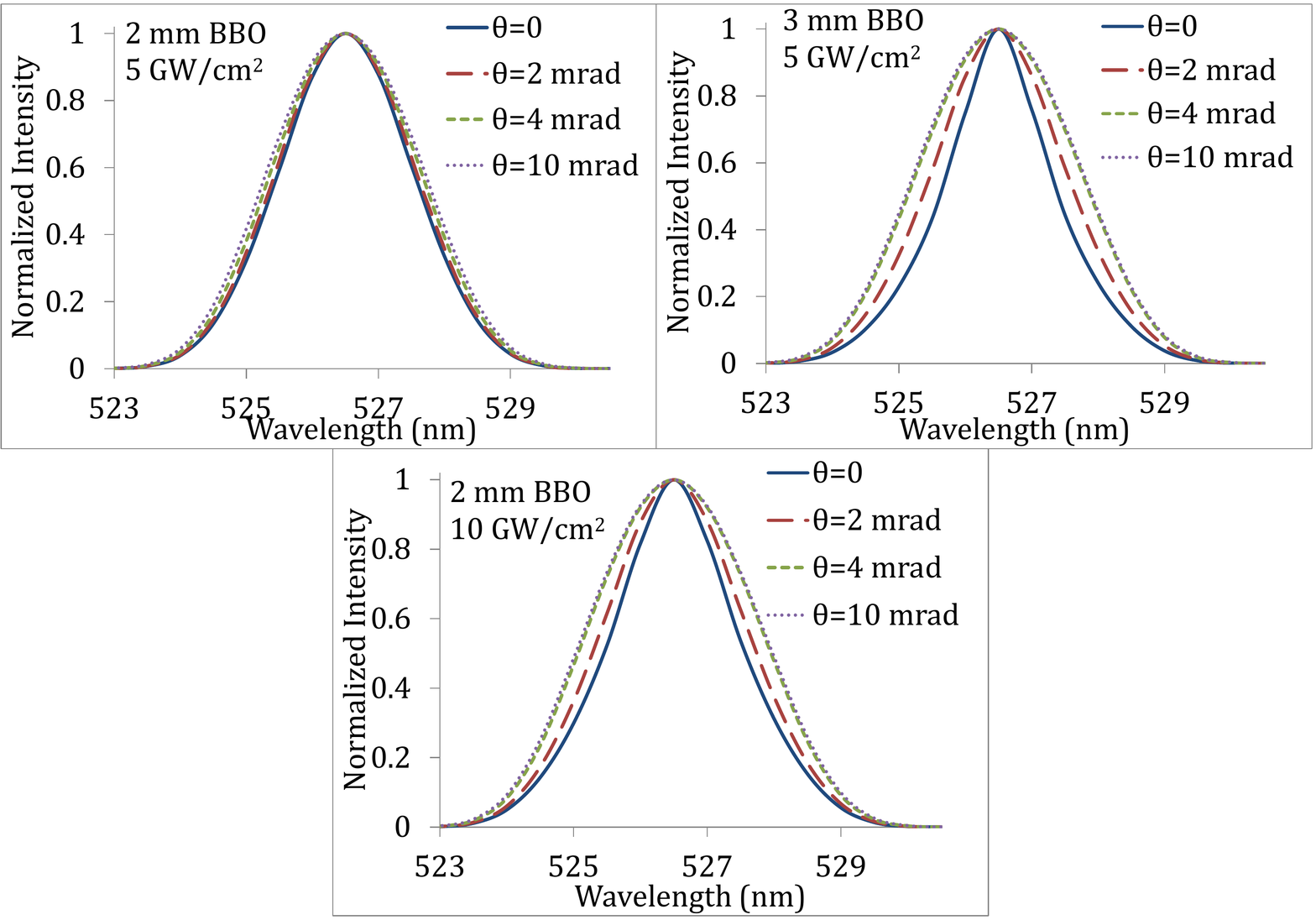}
\caption{A: Spectra of SHG pulse which are generated from pulses of 500 fs and 6.2 nm bandwidth and different beam divergence, pump intensities are 5 and 10 GW/cm$^{2}$.}
\label{fig7}
\end{figure}

Although the divergence of the input beam weakens the overall conversion efficiency, it helps preserving the bandwidth of the generated SH pulse for distances longer than the optimum length of the crystal. Figure(\ref{fig7}) shows spectra of SHG pulses which are generated from plane-wave and divergent beams at two thicknesses of the crystal and for two levels of the input intensity. The reason of this relationship between the beam divergence and the bandwidth of the SHG pulse can be explained  by examining the configuration of the nonlinear interaction of divergent beams and the dependence function of the conversion efficiency on the phase mismatch angle, as done in the following.

\begin{figure}[h!]
\centering
\includegraphics[ height=8cm, width=10cm]{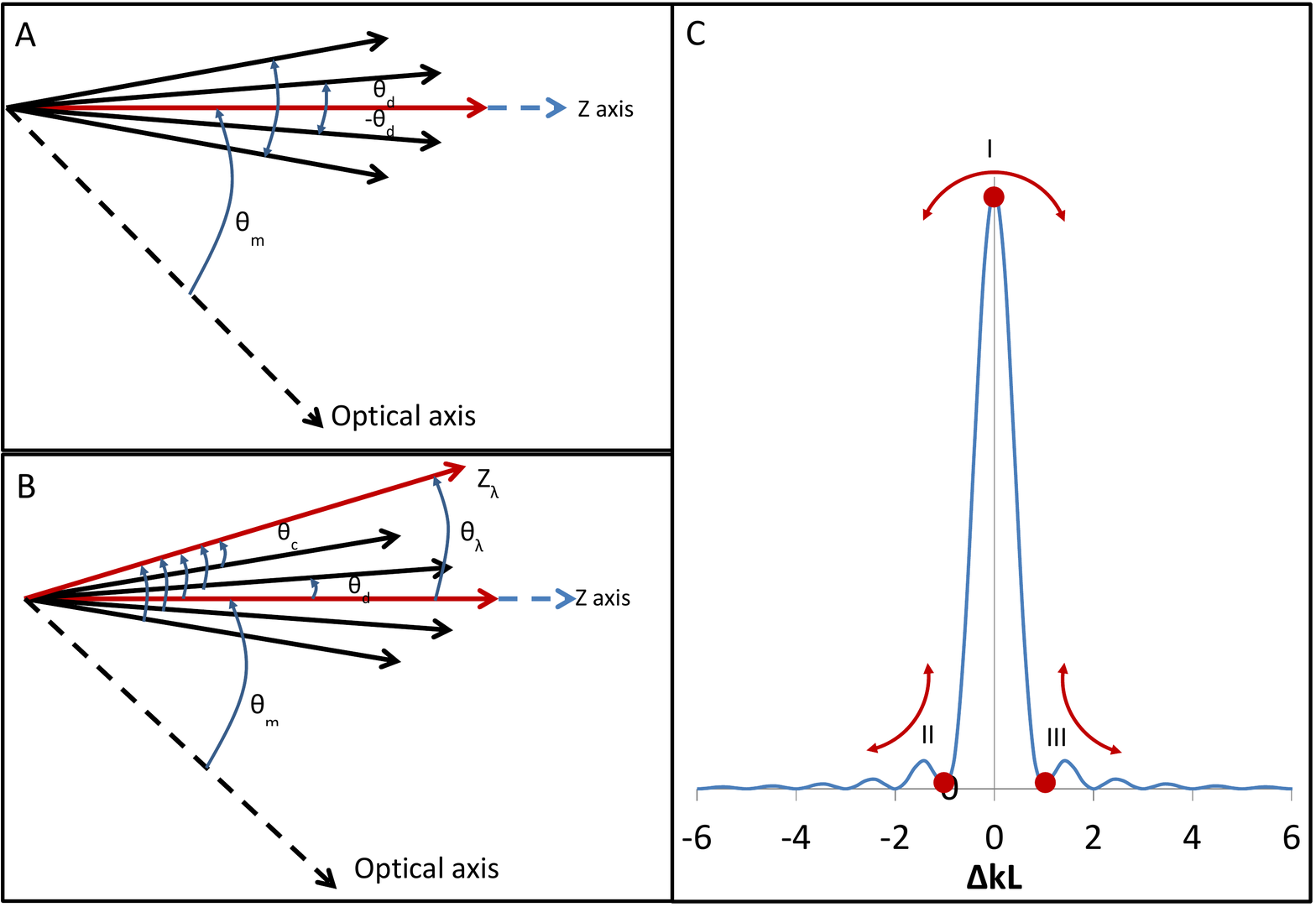}
\caption{A and B: the configurations and the phase mismatch angles of a monochromatic beam and a beam of finite bandwidth respectively, C: the phase matching tuning curve of the SHG. }
\label{fig8}
\end{figure}

As shown in figure(\ref{fig8} A), a monochromatic divergent beam is considered as a summation of plane-wave components that distribute around the beam axis with slightly different propagation angles $\mp \theta_{d} $. The central component is supposed to enter the crystal at the perfect phase matching angle $\theta_{m}$ with the crystal optical axis. In this case, $\theta_{d}$ is the deviation angle of each plane-wave component, which will, in combination with another FF plane-wave component, result in producing a SH plane-wave component that is deviated from the perfect phase matching direction. This results in a reduction of the conversion efficiency of the particular plane-wave component and consequently the overall efficiency. On the other hand, for a beam of a finite bandwidth, the phase matching angle is calculated only for the central wavelength. Thus, when the incident beam is assumed to be a plane wave, or to have a very small divergence, the non-central spectral components will have phase mismatch angles $\theta _{\lambda }$, figure(\ref{fig8} B), that result in degrading the  conversion efficiency of these components. 

In the plane-wave approximation, the phase matching tuning curve is a function of the wave vector mismatch and the crystal length, as shown in figure(\ref{fig8} C), which has FWHM and wings structure that depend on the input intensity \cite{han2008phase}. In the case, where the central spectral component enters the crystal at the perfect phase  matching angle, $\Delta k$ is a function of $\theta _{\lambda }$. Accordingly, the conversion efficiency of each spectral component can be represented by a point on the tuning curve in the frame C of the figure, where the efficiency of the central wavelength is at the peak of the curve and that of the other spectral components distribute around the peak.

\begin{figure}[h!]
\centering
\includegraphics[ height=8cm, width=10cm]{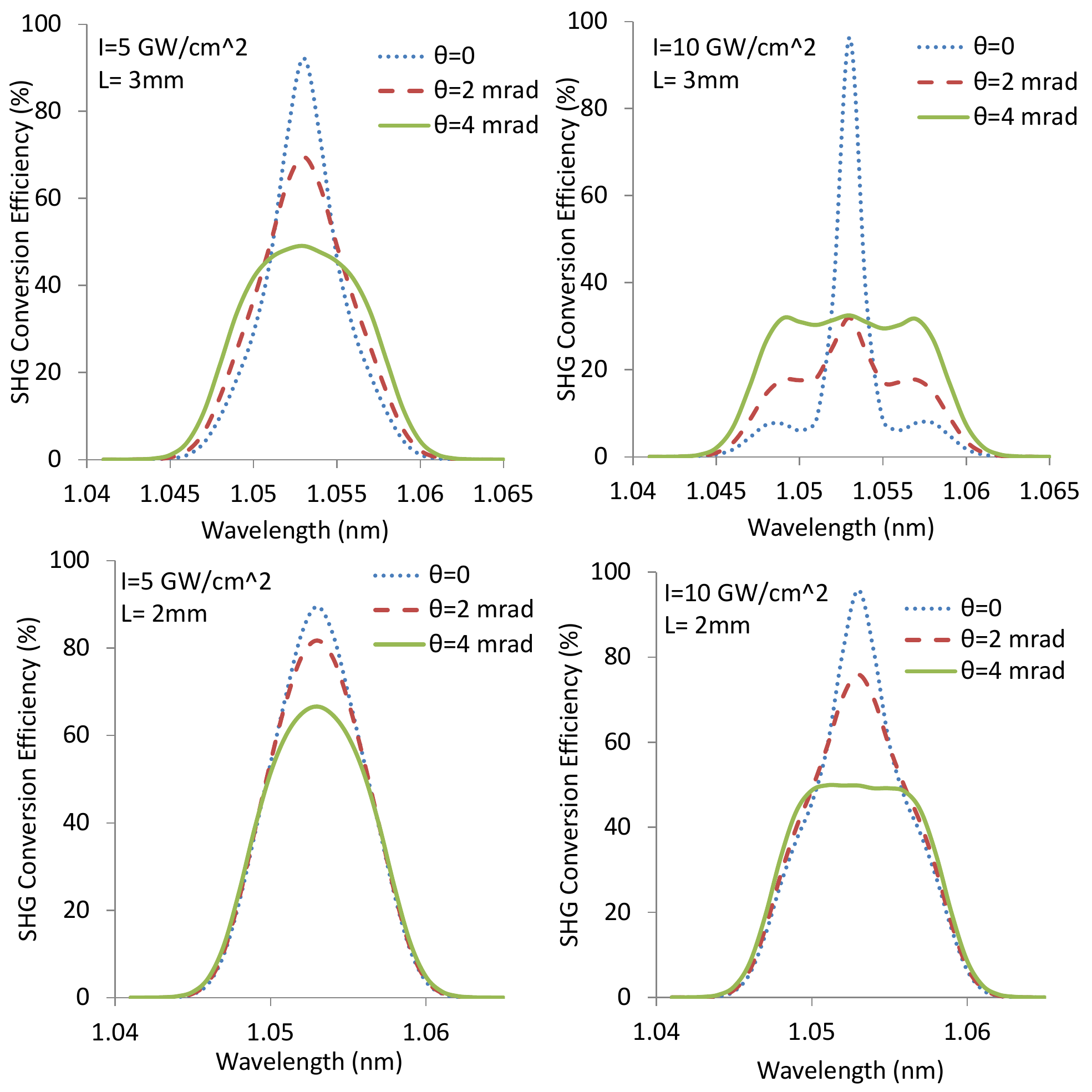}
\caption{The spectral acceptance function of the SHG process for two pump intensities and at two thicknesses of the BBO crystal for 500 fs input pulse}
\label{fig9}
\end{figure}

When the beam has a non-negligible divergence, the deviation angle of each plane-wave component for each spectral mode will be redefined as the angle between the propagation direction of the plane-wave and the phase matching direction of the particular spectral mode, $\theta _{c}$ in figure (\ref{fig8} B). For the central spectral component, where Z-axis is the phase matching direction, the plane-wave components on both sides of the phase matching direction will lead to reduce the conversion efficiency of this spectral component, because it is at the peak of the tuning curve, point I in the frame C, and the curve is symmetric at this point. For the non-central spectral components, where the phase matching directions are already off the Z-axis, e.g. the component $ Z_{\lambda} $, the plane-wave components which are above the Z-axis will help reducing the phase mismatching angles, while those under the Z-axis will do the opposite. Although the plane waves distribute symmetrically around the Z-axis, their influence on the conversion efficiency is not the same for the non-central spectral components, e.g. points II and III in the frame C of the figure. This is because the tuning curve at these regions is not symmetric, i.e. the enhancement of the efficiency toward the peak of the curve can be larger than the degradation in the other direction. Generally, the amount of the enhancement or the degradation of the conversion efficiency of a particular spectral component due to the beam divergence depends on the position of this component on the tuning curve. Consequently, this leads to change the shape of the spectral acceptance function of the crystal, resulting in producing SH pulses of different bandwidths from beams that have the same input bandwidth but have different divergence angles. Figure(\ref{fig9})presents examples for the spectral acceptance function at two levels of the input intensity and for two thicknesses of the non-linear crystal for 500 fs input pulse.

This function, besides the dependence on the divergence of the beam and the crystal properties, it depends on the input intensity and change with the crystal length, as shown in figure (\ref{fig9}). On the other hand, increasing the beam divergence above an optimum value, which is defined by the input pulse characteristics, will not lead to an enhancement in the bandwidth of the SHG. This can be seen clearly in figure (\ref{fig7}), where there is no spectral difference between the SHG pulses which are generated from beams of 4 mrad and 10 mrad divergence, where 4 mrad is the optimum beam divergence for that input bandwidth.

\begin{figure}[h!]
\centering
\includegraphics[ height=5cm, width=6cm]{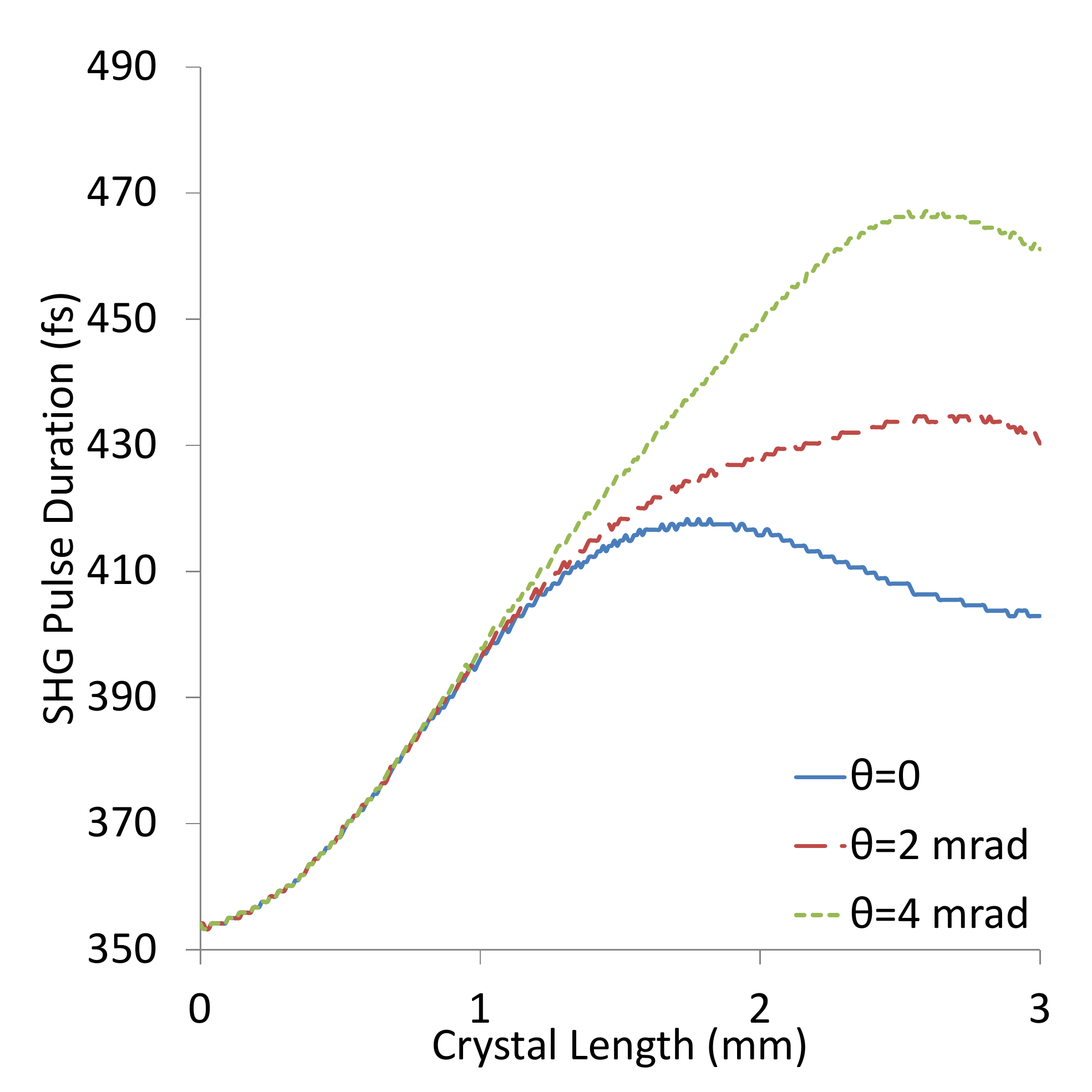}
\caption{SHG pulse duration as a function of the crystal length, input pulses are of different divergence all of 500 fs duration, 6.2 nm bandwidth and 5 GW/cm$^{2}$ intensity. }
\label{fig10}
\end{figure}

The consequence of the bandwidth preservation, in case of chirped pulses, is that the SHG pulse, that is generated from divergent beam, continues being broadened due to the GVM as the pulses propagate into the crystal, resulting in a longer SHG pulse, as shown in figure(\ref{fig10}). 

\subsection{Optical Parametric Amplification}

OPA is a second order nonlinear process in which a high intensity pump beam amplifies a low intensity seed having lower frequency. In addition an idler beam is generated at the frequency that satisfies the energy conservation law. In a degenerate OPA the pump beam is at the second harmonic frequency of the seed and the idler, which are at the fundamental harmonic.

The nonlinear interaction of pump, seed and idler at perfect phase matching in a nonlinear medium consists of cycles of sequential difference and sum frequency generation processes \cite{richard1979tica}. After the pump is depleted in a difference frequency process, a sum frequency process starts to create pump photons at the expense of the signal and the idler photons. For the same input wavelength and in the same crystal, the conversion length, which is the crystal length over which the pump is fully depleted, is a function of the total input intensity. At constant pump intensity, the conversion length is a function of the seed intensity. 

However, in short pulse OPAs, GVM between the pump and the signal/ idler derails the sequence of the difference and sum frequency processes by changing the local overlapped intensities as the pulses propagate along the crystal. Besides limiting the system efficiency, the interplay of the pump depletion and the GVM, can lead to significant modifications in the spatial and temporal profiles of the resulting pulses. Accordingly, in low gain regime, for optimum efficiency and optimum pulse characteristics, the thickness of the crystal should be carefully selected to fit the optimum conversion length of the incident seed and pump pulses. For fine optimization in a particular crystal, the seed energy can be tuned, or simply a small delay can be introduced between the pump and the seed, as will be shown later.

The distortions on the temporal and the spatial profiles of the generated pulses are related to the dependence of the non-linear process direction, up or down conversion, on the local overlapped intensities, i.e. some parts of the signal and the idler pulses may be undergoing amplification while other parts are having up-conversion interaction. In this case and under strong pump depletion the temporal profiles of the signal and the idler can have asymmetric modifications around the peaks of the pulses. In the same way the generated beams can suffer spatial degradation, which is symmetric around the beam axis when the crystal is much shorter than the spatial walk-off length of the pump.

    When the pump and the seed are divergent beams, certain parts of the beams do not satisfy the perfect phase matching condition. The effect of the dephasing is that the parametric process is terminated before reaching the point of full pump depletion, resulting in a reduction of  the maximum conversion efficiency, after which the process starts again without reaching the point of zero gain. The deviation of the processes from the perfect sequence increases as the divergence of the beams increases. However, the results of the simulation show that in the short pulse regime, where focusing the beams is not required, OPA efficiency is not as sensitive to the beam divergence such as it is for the SHG process.

In the simulation of the OPA, we consider the output of 3 mm BBO in the SHG stage as the pump of the OPA. The seed is considered, arbitrarily, as 6$\%$ of the input energy of the system. The simulation allows to introduce pre-delay between the seed and the pump for the fine optimization of the idler characteristics and the system efficiency. The calculations are for TL and linearly chirped pulses. In case of non-TL pulses, the pump pulses were shorter than the seed due to the reduction of the pulse duration in the SHG process.
\subsubsection{Transform limited input pulse}
\begin{figure}[h!]
\centering\includegraphics[height=5cm, width=5cm]{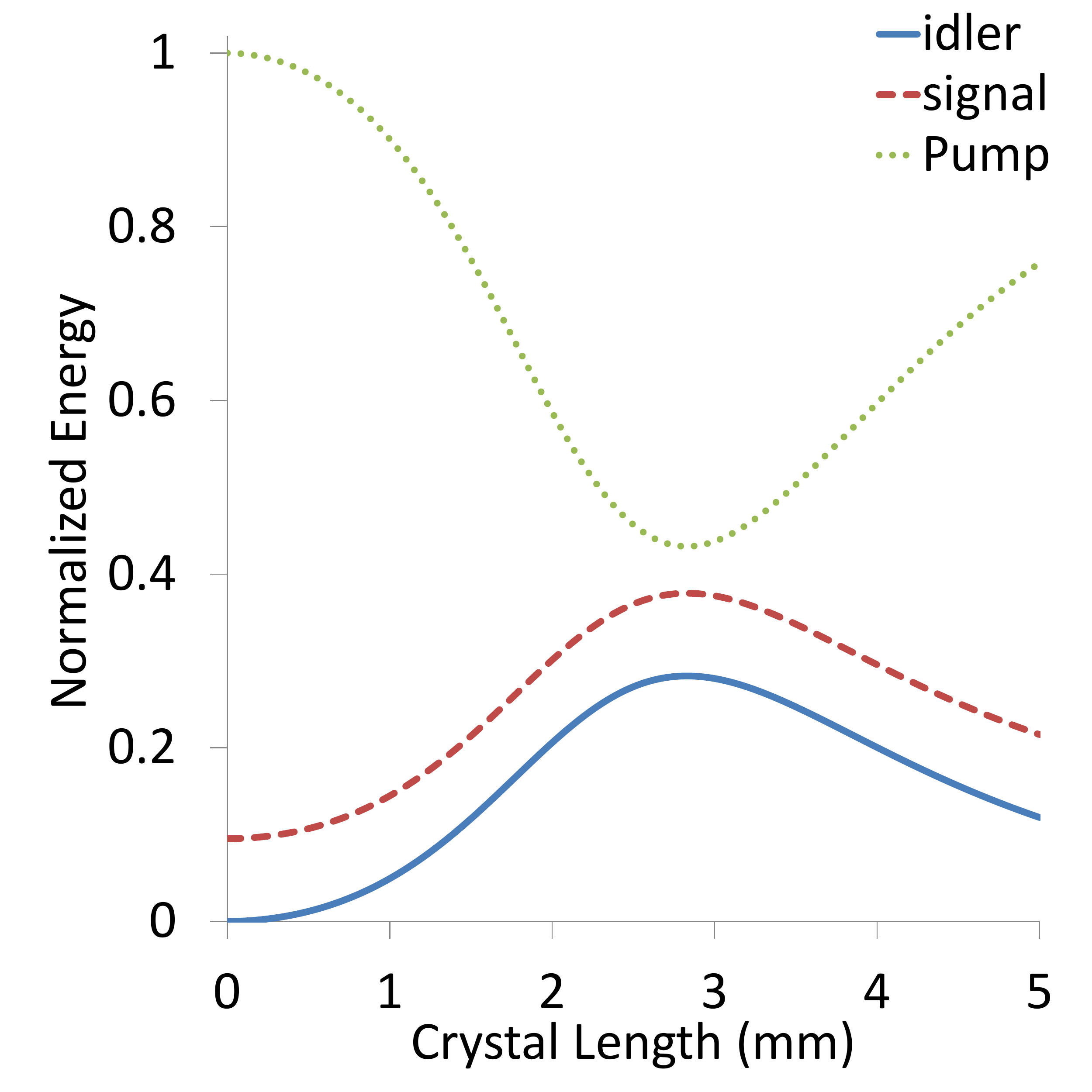}
\caption{ Normalized energies of pump, signal and idler as functions of the crystal length, total input intensity to the system is 5 GW/cm $^{2}$, the seed pulse duration is 500 fs, the pump is 480 fs and the external non-collinear angle is 0.6$^{o}$.}
\label{fig11}
\end{figure}

Figure (\ref{fig11}) shows an example of pump, signal and idler normalized energies behaviour, in non-collinear degenerate OPA, as function of the crystal length, when the input pulse to the system is TL of duration 500 fs and intensity 5 GW/cm$^{2}$. In this case the seed and the pump are perfectly synchronized in time and mixed at external angle 0.6$ ^{o} $. As shown, for this level of the input intensities, the point of full pump depletion is not applicable because of the GVM between the pump and the signal/idler.

When this interaction is performed in practice, it is unlikely that a crystal of realistic thickness will exactly fit the optimum conversion length, i.e. the thickness can not be tuned. Thus, a crystal slightly shorter than the optimum length (eg. 2 mm in this case) can be used and an idler pulse that is free from any degradation can be obtained, but this is done at the expense of the process efficiency. However, a degradation-free idler pulse can also be produced by fine tuning the interacting pulses to fit a crystal of thickness slightly longer than the optimum length. This can be done by introducing a small, suitable delay between the seed and the pump to obtain the optimum conversion at the exit of the crystal, which can be done, in this case, by delaying the seed. By using this technique, the spatial, temporal and spectral profiles of the idler can be optimized and, equally important, the efficiency of the stage can be maximized up to a value that can be even higher than that at perfect synchronization. In addition , when the input pulse is non-TL, introducing delay is imperative for the idler frequency shift correction, as will be discussed later.

\begin{figure}[h!]
\centering
\includegraphics[ height=5cm, width=10cm]{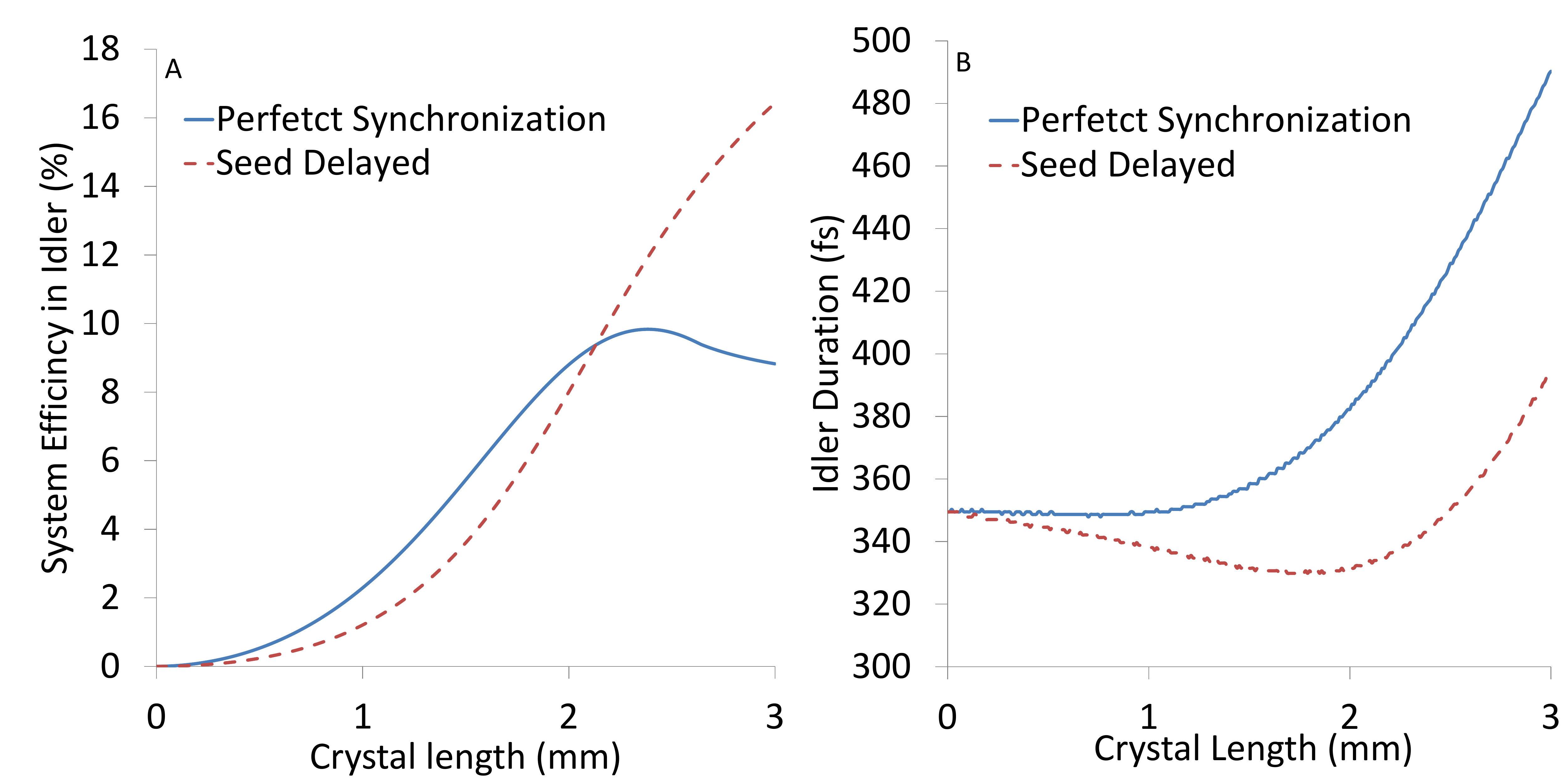}
\caption{A: The energy conversion efficiency of the system, idler energy to the total input energy to the system, for TL pulse of 500 fs duration and 5 GW/cm$ ^{2} $ intensity at the system input when the seed and the pump are perfectly synchronized and when the seed is delayed, B: the variation of the idler pulse duration along the crystal in both cases.}
\label{fig12}
\end{figure}

In this case, where $v_{gs} > v_{gp}$, delaying the seed means that this will scan the pump pulse as both pulses propagate into the crystal and then will gain more energy instead of walking away from the pump after being perfectly matched at the crystal input and the same applies to the idler. The other consequence of delaying the seed is that the effect of GVM in broadening the signal and the idler pulses is reduced. Figures ((\ref{fig12}) A) present the energy conversion efficiency of the whole system, in the idler, for the two cases of perfect synchronization and when the seed is delayed in order to make the conversion length fit 3 mm crystal. The variation of the generated idler pulse duration along the crystal for both cases in figures ((\ref{fig12}) A) are shown in frame B of the figure. The total input energy and the pulse duration are the same as for figure(\ref{fig11}).   

 For a particular TL input pulse the amount of the optimum delay is defined by two factors: (a) the difference between the conversion length at perfect synchronization and the thickness of the chosen crystal, (b) the difference between the seed and pump group velocities along the pump propagation direction, which is, for a particular input pulse, a function of the non-collinear angle $ (\Omega ) $, since the velocity difference in this case is ($ v_{gs} \cos(\Omega ) - v_{gp} $).
 
 The enhancement of the intensity-transfer characteristics in OPAs by introducing suitable delay between the seed and the pump has been studied in \cite{wang1994optical}. They concluded that, when the group velocity mismatched SHG technique is used to generate a short pump pulse, the amount of the required delay in the OPA stage almost does not depend on the seed intensity. However, the case is different for low gain OPA, where the seed intensity is an effective parameter on the conversion length of the OPA, i.e. in a particular crystal, changing the seed intensity effectively changes the conversion length of the OPA process, consequently, changing the seed intensity changes the required delay for optimization. In addition, when the input pulse is a non-TL pulses, the spectral shift of the idler is another factor that defines the required delay in the OPA, as will be shown later.

\begin{figure}[h!]
\centering
\includegraphics[ height=6cm, width=12.65cm]{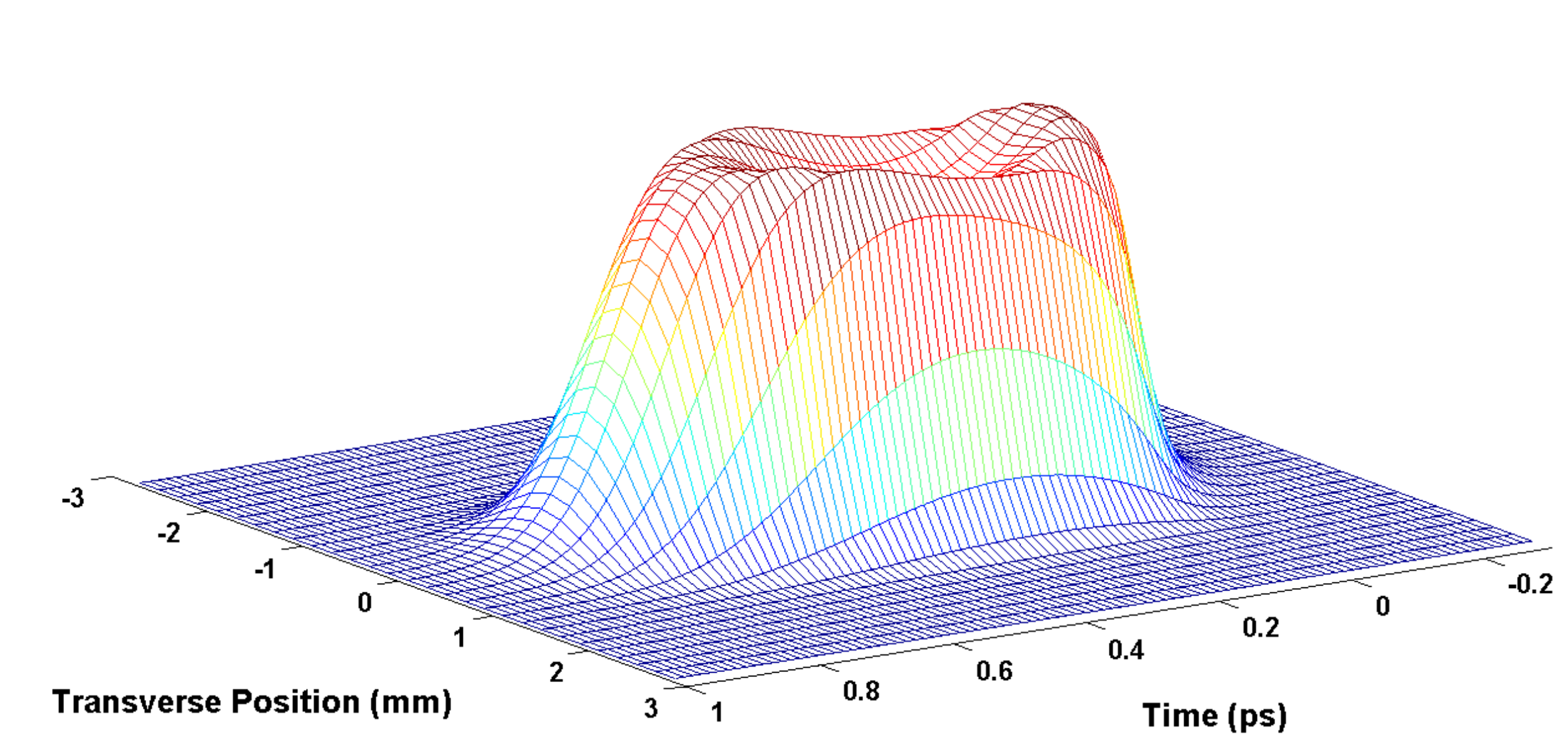}
\caption{Idler pulse resulting from 4 mm crystal in time and space, the  input pulse to the system is TL of 500 fs duration and 5 GW/cm$ ^{2} $ intensity.}
\label{fig13}
\end{figure}

\begin{figure}
\centering
\includegraphics[ height=5cm, width=7cm]{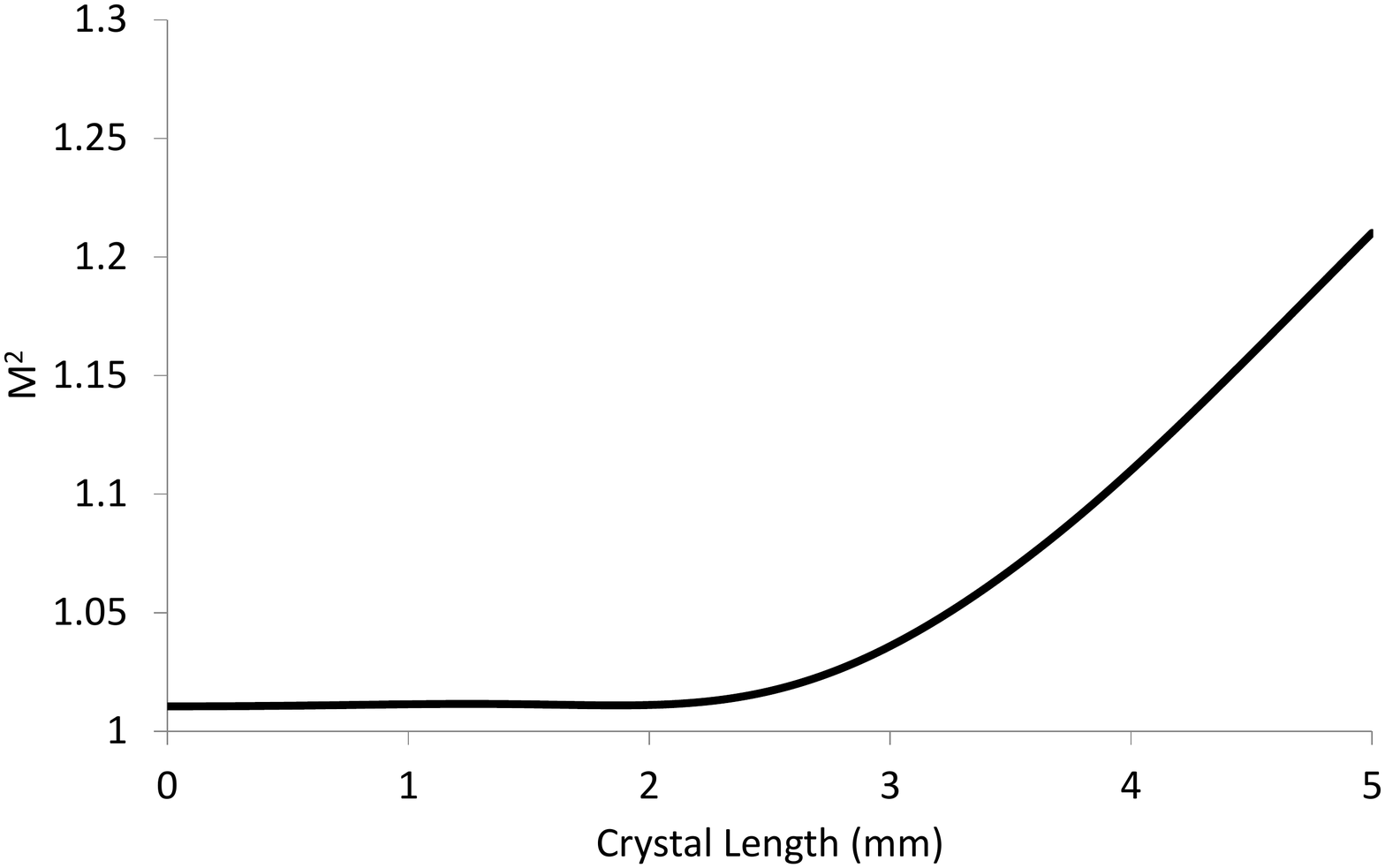}
\caption{ M$^{2}$ values as a function of the crystal length of idler resulting from 500 fs pulse of 6 GW/cm$ ^{2} $ intensity at the system input .}
\label{fig14}
\end{figure}

On the other hand, achieving maximum idler intensity or energy is not the only parameter that defines the optimum delay, but also the temporal, the spatial and the spectral profiles of the idler should be considered. Using a thick crystal, for example 4 or 5 mm for the pulse under study, produces an idler pulse that is deformed in the space, time and spectral domains. Figure (\ref{fig13}) presents the deformation of the idler pulse in time and space when a 4 mm crystal is used in the OPA stage. Figure (\ref{fig14}) shows the M$ ^{2} $ value of the generated idler beam along a 5 mm BBO crystal. 

\begin{figure}[h!]
\centering
\includegraphics[ height=7cm, width=11.66cm]{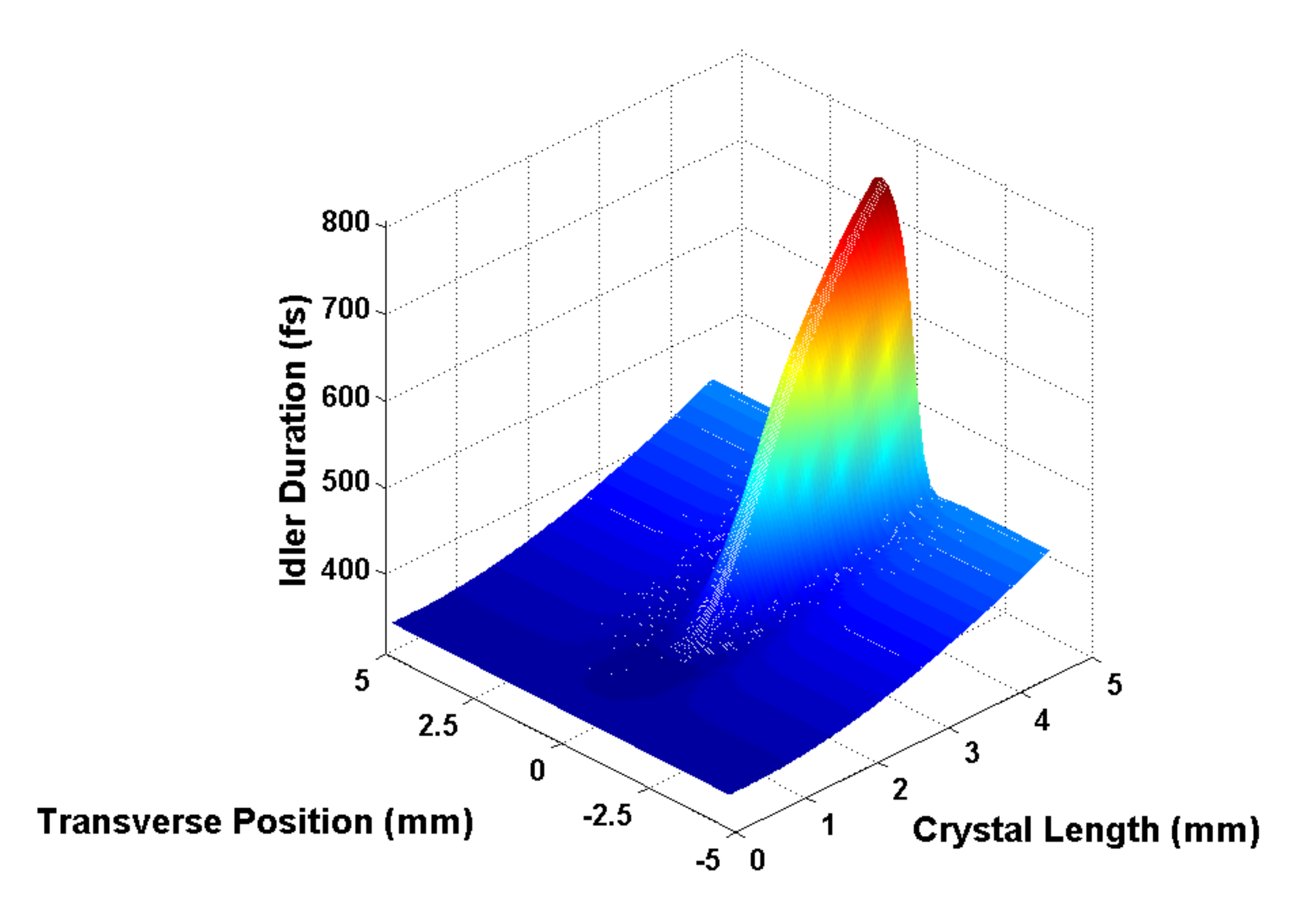}
\caption{Pulse duration across the beam as a function of the crystal length of idler resulting from 500 fs pulse of 6 GW/cm$ ^{2} $ intensity at the system input .}
\label{fig15}
\end{figure}

Moreover, beyond the optimum crystal length the resulting idler pulse can have different pulse duration across the beam, as shown in figure (\ref{fig15}). This happens since the interplay of the GVM and the process saturation affects the resulting pulse differently in space depending on the local intensity.

\subsubsection{Non-transform limited input pulse}
\begin{figure}[h!]
\centering
\includegraphics[ height=9 cm, width=9cm]{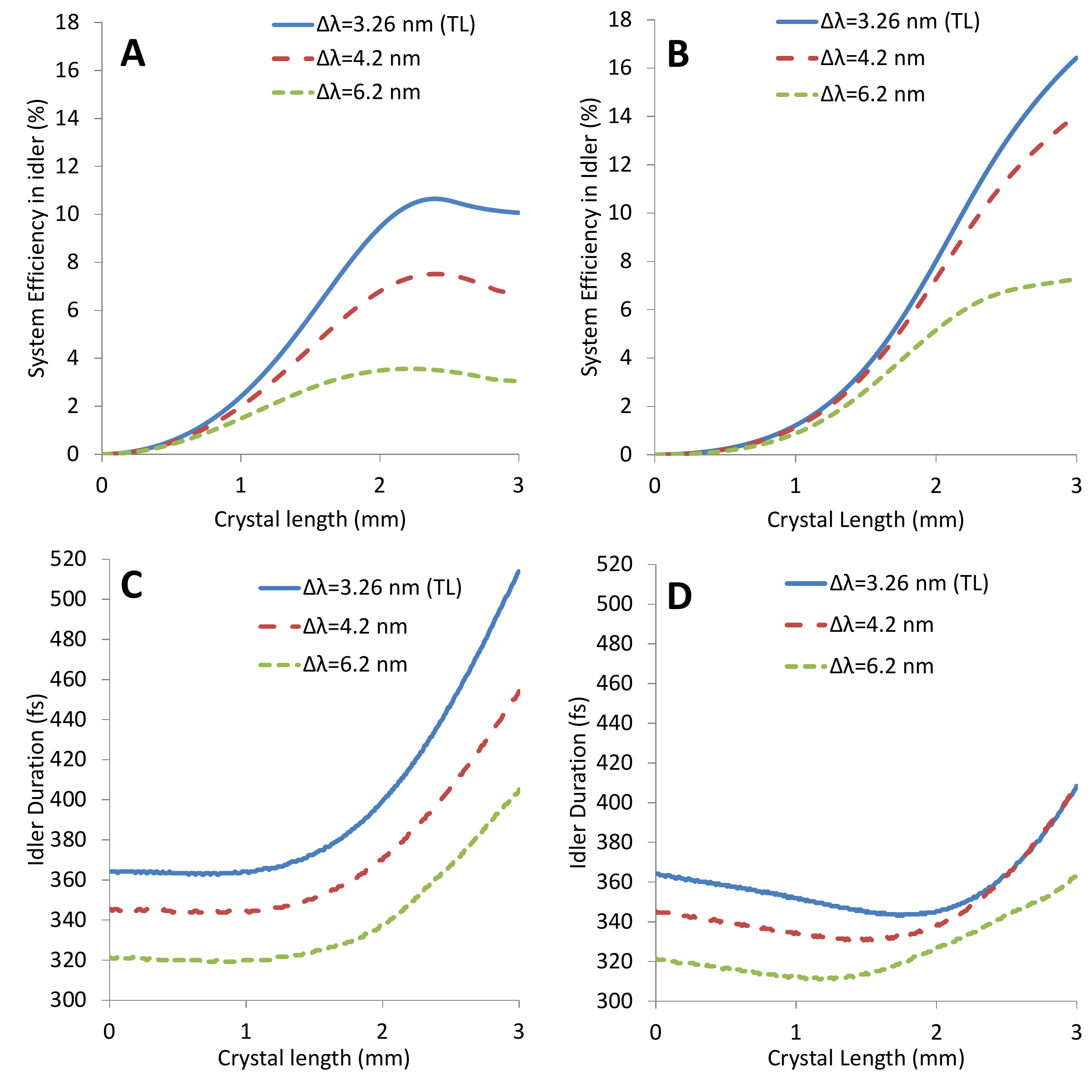}
\caption{A and B The overall energy efficiency of the system for pulses of 500 fs duration and 5 GW/cm$ ^{2} $ intensity but of different spectral bandwidths in case of perfect synchronization and when the seed is delayed respectively, C and D the duration of the resulting idler along the crystal for the cases A and B respectively, the non-collinear angle is 0.6$ ^{o} $.}
\label{fig16}
\end{figure}

When the input pulse is a non-TL pulse, the maximum overall efficiency of the system is reduced because of the dependence of the two non-linear processes on the spectral content of the input pulse. Figures (\ref{fig16} A and B) show examples of the overall energy conversion efficiency of the system, in the idler, for pulses of three different spectral bandwidths but having the same input intensity and pulse duration.

In addition, the bandwidth of the input pulse determines the duration of the generated idler pulse through three effects. First, as mentioned earlier, the duration of the generated SH pulse depends on the input pulse bandwidth, which will, then, define the gain window in the OPA stage and then the idler duration. Second, the dependence of the SHG efficiency on the input pulse bandwidth leads to producing different seed-pump intensity ratios for different bandwidths of the input pulse, which, thereafter, need different seed-pump delay in the OPA stage in order to optimize the efficiency, which leads to  generating of idler pulses having different pulse durations. Finally, even at fixed seed/pump intensity ratio and pulses durations, the optimum delay in the OPA depends on the chirp rate of the the pulses, because of the dependence of the conversion efficiency on the spectral content of the pulses. This leads to generating of idler pulses having different durations from input pulses of the same initial duration but of different bandwidths, as shown in frames C and D of figure (\ref{fig16}) for the cases A and B respectively in the same figure.   

The simulation shows that the required delay for simultaneously optimizing   the conversion efficiency and the idler characteristics is the delay that maximises the summation of the time and the frequency integrations of the generated idler pulse.

 \subsubsection{The spectral profile of the generated idler}
\begin{figure}
\centering
\includegraphics[ height=5 cm, width=12 cm]{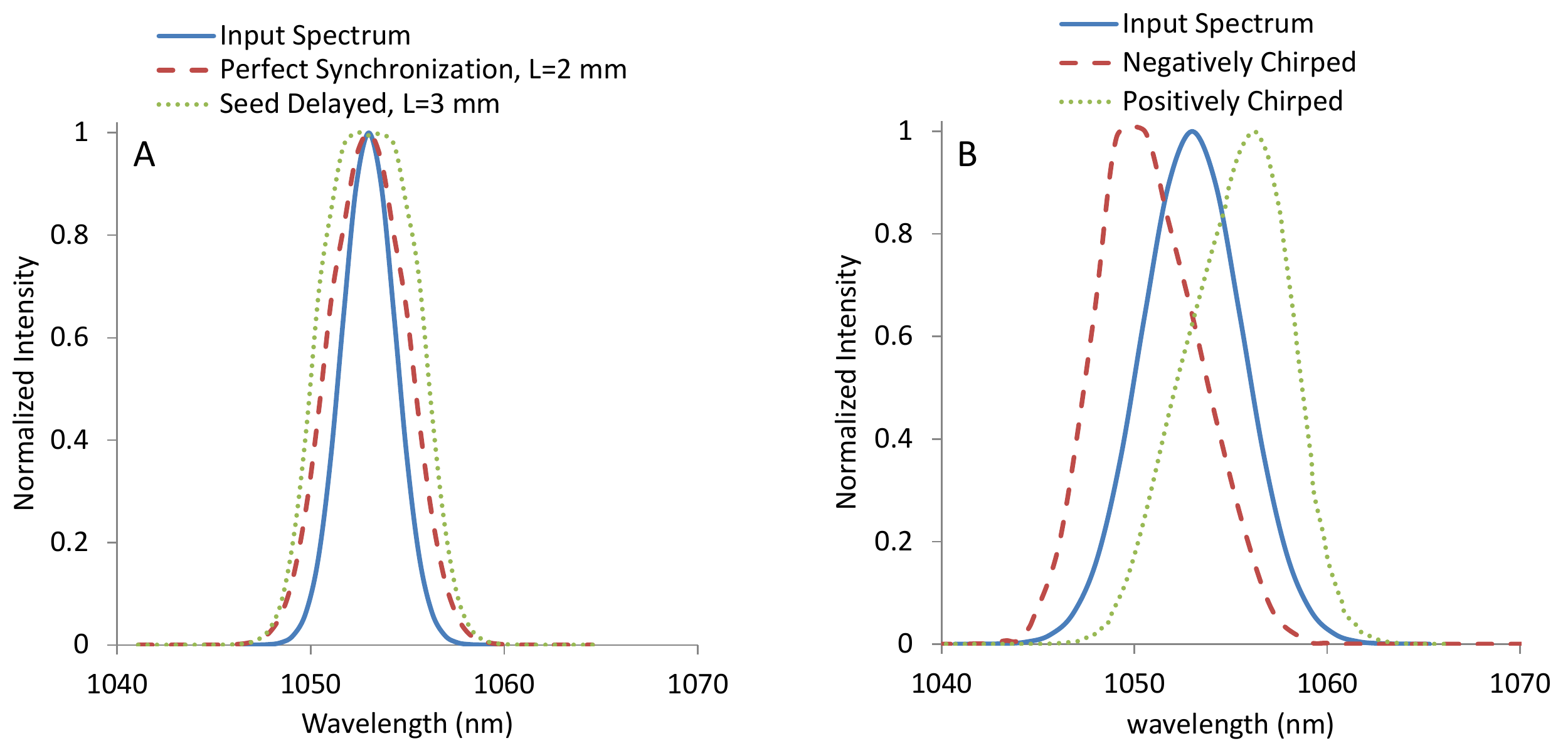}
\caption{The spectra of idler pulse which is produced from 500 fs pulse of 5 GW/cm$ ^{2} $ intensity at 0.6$ ^{o} $ non-collinear angle when (A) the input pulse is TL, L is the crystal length (B) the input pulse having 6.2 nm bandwidth.}
\label{fig17}
\end{figure}

Working with a degenerate OPA, the bandwidth of the beam can be preserved throughout the system, due to the broadband amplification ability of the OPA in this regime. When the input pulse is TL, the generated idler pulse spectrally can be broader than the input pulse without any frequency shift when a crystal of an optimum length is used. In addition, delaying the faster pulse, for efficiency optimization, in a crystal that is slightly longer than the optimum thickness does not cause any reduction in the spectrum of the resulting idler, but, on the contrary, slightly enhances the idler bandwidth. Figure (\ref{fig17} A) shows two examples of the resulting idler spectra when the pump and the seed are perfectly synchronized at the input of 2 mm crystal, and when the seed is delayed to match the conversion length to a 3 mm crystal. 

On the other hand, when the input pulse is chirped, the spectrum of the generated idler is shifted from the input spectrum due to GVM between the signal/idler and the pump. The shift direction depends on the chirp sign of the input pulse, blue shift when the input pulse is  negatively chirped and red shift for the positively chirped, as shown in figure (\ref{fig17} B). In low-gain regime, the amount of the spectral shift depends on the initial chirp of the input pulse and the saturation level of the process, which is dependent upon the seed/pump intensity ratio and the crystal length.

The influence of GVM, in this case, can be eliminated by mixing the seed and the pump at an angle at which $v_{gs} \cos(\Omega)= v_{gp}$. However, this leads to generated idler beam with large angular dispersion, as will be discussed later. Another way for bringing back the spectrum of the idler to match the input spectrum is delaying the seed pulse relative to the pump. The required delay is defined by the GVM of the pulses in the OPA crystal, the chirp rate of the input pulse and the difference between the conversion length of the OPA and the thickness of the crystal. According to the numerical simulation, the required delay for optimizing the conversion efficiency in a crystal that is a bit longer than the conversion length is equivalent to that which is required for bringing the spectrum to the desired position. Thus, using a crystal that is slightly longer than the conversion length and delay the faster pulse ensure optimum efficiency for the process and eliminate the spectral shift in the idler at the same time.       
\subsubsection{The spatial profile of the generated idler}
\begin{figure}[h!]
\centering
\includegraphics[ height=5 cm, width=6cm]{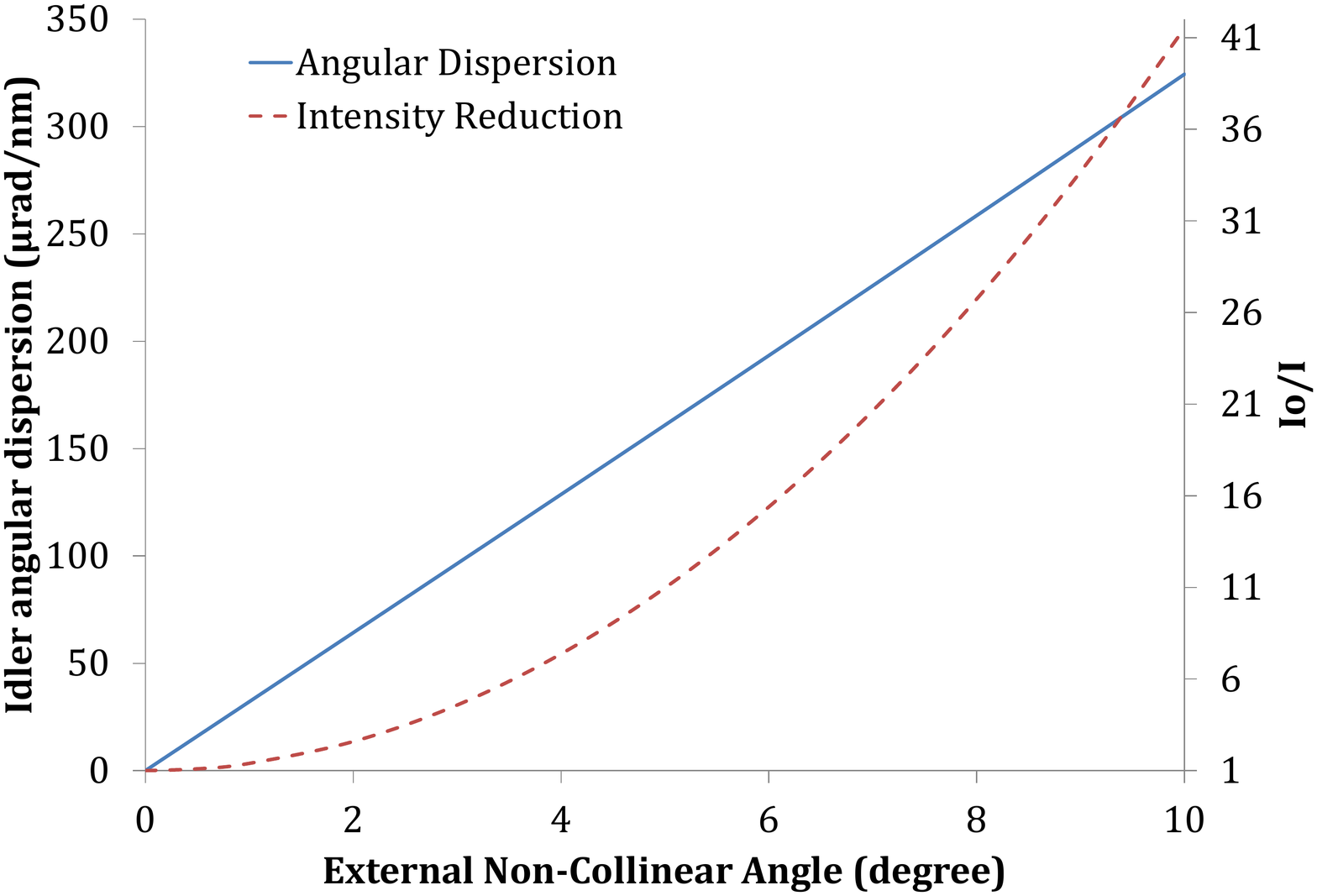}
\caption{The angular dispersion of the idler beam and the total intensity reduction in the focal plane as a function of the non-collinear angle for idler beam having 5 nm bandwidth, 500 fs duration and 3 mm beam diameter, I, I$ _{o} $ the intensities in the focus spot for beams with and without angular respectively.}
\label{fig18}
\end{figure}

At optimum working conditions of the system, the idler beam radius is about 70-80$\%$ of the input beam radius depending on the saturation level of the processes. On the other hand, in such a system, when type I phase matching is used, OPA needs to work in the non-collinear geometry in order to separate the resulting idler beam from the signal, because the two beams have the same wavelengths and polarization. Despite the advantages of the non-collinear interaction geometry, such as the broadband amplification and group velocity matching, the resulting idler beam from this configuration is angularly dispersed in the phase matching plane of the OPA. Since the idler will be seeded to a high power laser system, the angular dispersion, and hence the front tilt, will affect the performance of the system. For instance, the angular dispersion and the accompanying  front tilt lead to an enlargement of the focal spot and broadening of the pulse duration \cite{pretzler2000angular}. Figure (\ref{fig18}) shows the angular dispersion of the generated idler, assuming narrow bandwidth, and the ratio of the detected intensity in the focal plane, by a detector perpendicular to the propagation direction of the beam, to the intensity of a beam without angular dispersion as a function of the non-collinear angle. 

Accordingly, since in the degenerate regime, the broadband amplification is already achievable and the pulse bandwidth, in the range of the study, is narrow enough to be amplified without optimizing the non-collinear angle, mixing the seed and the pump at small non-collinear angle reduces the effect of idler angular dispersion down to a negligible value, where the spatial and the temporal changes can be ignored. Otherwise, pulse-front tilted technique \cite{huang2012broadband} or post amplification dispersion correction technique \cite{shirakawa1998pulse} should be used in order to compensate for the angular dispersion of the idler beam.

\section{Conclusions}
We have described a reliable model for characterizing and optimizing a second harmonic generation-optical parametric amplification based temporal contrast enhancement system. The model assumes a common source for the second harmonic generation and the optical parametric amplification and simultaneously takes into account the group velocity mismatch, the beam divergence, the pump depletion, and the spectral content of the input pulse for transform and non-transform limited pulses. Numerical solutions of the model predict the performance of both the related nonlinear processes and give clear evidence of the dependence of the system performance on the input pulse characterization and the system configuration. The group velocity mismatch, the pump depletion, the crystal length, and the spectral content of the input pulse are the most effective parameters that affect the performance of the optical parametric stage. In addition, the beam divergence determines the efficiency, the temporal and the spectral shapes of the resulting pulse in the second harmonic generation stage. A criterion has been set to show the way in which the beam divergence affects the efficiency of the second harmonic generation by defining a length that represents the coherence length at pump depletion regime.  The efficiency of the system and the idler characteristics is efficiently optimized by using a crystal slightly longer than the conversion length and introducing small delay between the seed and the pump. Using this way the spectral shift of the idler, in case of non-transform limited input pulse, can be corrected simultaneously. The efficiency and the temporal shape of the generated idler pulse depend on the spectral content of the input pulse and on the seed-pump synchronization. The study provides a comprehensive understanding for the temporal contrast enhancement unit and enables implementing the enhancement in a way that ensures generating a high contrast and high quality seed for a high power laser system.\\

\textbf{Acknowledgements}\\

We acknowledge the  financial support from EPSRC (grant numbers EP/I029206/1 and EP/L013975/1) and the Iraqi Ministry of Higher Education and Scientific Research. We also acknowledge the assistance of the Center of Plasma Physics staff at Queen's University Belfast.



\end{document}